\documentclass[12pt,english]{article}
\usepackage{times}
\usepackage[T1]{fontenc}
\usepackage{geometry}
\usepackage{graphicx}
\usepackage{setspace}
\usepackage{amssymb}

\makeatletter

\providecommand{\tabularnewline}{\\}

\usepackage{bm}

\usepackage{babel}
\makeatother
\begin{document}

\section*{\noindent Multi-Functional Polarization-Based Coverage Control through
Static Passive EMSs}

\noindent ~

\noindent \vfill

\noindent G. Oliveri,$^{(1)(2)}$ \emph{Fellow}, \emph{IEEE}, F. Zardi,$^{(1)(2)}$
\emph{Member}, \emph{IEEE}, A. Salas-Sanchez,$^{(1)(2)}$ \emph{Member},
\emph{IEEE}, and A. Massa,$^{(1)(2)(3)(4)(5)}$ \emph{Fellow, IEEE}

\noindent \vfill

\noindent ~

\noindent {\footnotesize $^{(1)}$} \emph{\footnotesize ELEDIA Research
Center} {\footnotesize (}\emph{\footnotesize ELEDIA}{\footnotesize @}\emph{\footnotesize UniTN}
{\footnotesize - University of Trento)}{\footnotesize \par}

\noindent {\footnotesize DICAM - Department of Civil, Environmental,
and Mechanical Engineering}{\footnotesize \par}

\noindent {\footnotesize Via Mesiano 77, 38123 Trento - Italy}{\footnotesize \par}

\noindent \textit{\emph{\footnotesize E-mail:}} {\footnotesize \{}\emph{\footnotesize giacomo.oliveri,
francesco.zardi, aaron.salassanchez@unitn.it, andrea.massa}{\footnotesize \}@}\emph{\footnotesize unitn.it}{\footnotesize \par}

\noindent {\footnotesize Website:} \emph{\footnotesize www.eledia.org/eledia-unitn}{\footnotesize \par}

\noindent {\footnotesize ~}{\footnotesize \par}

\noindent {\footnotesize $^{(2)}$} \emph{\footnotesize CNIT - \char`\"{}University
of Trento\char`\"{} ELEDIA Research Unit }{\footnotesize \par}

\noindent {\footnotesize Via Sommarive 9, 38123 Trento - Italy}{\footnotesize \par}

\noindent {\footnotesize Website:} \emph{\footnotesize www.eledia.org/eledia-unitn}{\footnotesize \par}

\noindent {\footnotesize ~}{\footnotesize \par}

\noindent {\footnotesize $^{(3)}$} \emph{\footnotesize ELEDIA Research
Center} {\footnotesize (}\emph{\footnotesize ELEDIA}{\footnotesize @}\emph{\footnotesize UESTC}
{\footnotesize - UESTC)}{\footnotesize \par}

\noindent {\footnotesize School of Electronic Science and Engineering,
Chengdu 611731 - China}{\footnotesize \par}

\noindent \textit{\emph{\footnotesize E-mail:}} \emph{\footnotesize andrea.massa@uestc.edu.cn}{\footnotesize \par}

\noindent {\footnotesize Website:} \emph{\footnotesize www.eledia.org/eledia}{\footnotesize -}\emph{\footnotesize uestc}{\footnotesize \par}

\noindent {\footnotesize ~}{\footnotesize \par}

\noindent {\footnotesize $^{(4)}$} \emph{\footnotesize ELEDIA Research
Center} {\footnotesize (}\emph{\footnotesize ELEDIA@TSINGHUA} {\footnotesize -
Tsinghua University)}{\footnotesize \par}

\noindent {\footnotesize 30 Shuangqing Rd, 100084 Haidian, Beijing
- China}{\footnotesize \par}

\noindent {\footnotesize E-mail:} \emph{\footnotesize andrea.massa@tsinghua.edu.cn}{\footnotesize \par}

\noindent {\footnotesize Website:} \emph{\footnotesize www.eledia.org/eledia-tsinghua}{\footnotesize \par}

\noindent {\small ~}{\small \par}

\noindent {\small $^{(5)}$} {\footnotesize School of Electrical Engineering}{\footnotesize \par}

\noindent {\footnotesize Tel Aviv University, Tel Aviv 69978 - Israel}{\footnotesize \par}

\noindent \textit{\emph{\footnotesize E-mail:}} \emph{\footnotesize andrea.massa@eng.tau.ac.il}{\footnotesize \par}

\noindent {\footnotesize Website:} \emph{\footnotesize https://engineering.tau.ac.il/}{\footnotesize \par}

\noindent \vfill

\noindent \emph{This work has been submitted to the IEEE for possible
publication. Copyright may be transferred without notice, after which
this version may no longer be accessible.}

\noindent \vfill

\newpage
\section*{Multi-Functional Polarization-Based Coverage Control through Static
Passive EMSs}

~

~

~

\begin{flushleft}G. Oliveri, F. Zardi, A. Salas-Sanchez, and A. Massa\end{flushleft}

\vfill

\begin{abstract}
\noindent An innovative multi-functional static-passive electromagnetic
skin (\emph{SP-EMS}) solution is proposed to simultaneously support,
in reflection, two independent wave-manipulation functionalities with
a single meta-atoms arrangement on the \emph{EMS} aperture when illuminated
by two \emph{EM} sources operating at the same frequency, but working
in different polarization states. Towards this end, a simple reference
meta-atom is designed first to enable an accurate and independent
control of each polarization component of the local reflection tensor.
Successively, the macro-scale \emph{}synthesis of multi-polarization
(\emph{MP}) \emph{SP-EMS}s (\emph{MP-SP-EMS}s) is carried out by solving
a global optimization problem where a cost function, which mathematically
codes separate requirements for each polarization, is minimized with
a customized version of the system-by-design (\emph{SbD}) technique.
Representative results from a set of numerical and experimental tests
are reported to assess the feasibility of a multi-function \emph{EMS}
based on polarization diversity as well as the effectiveness and the
robustness of the proposed method for the synthesis of \emph{MP-SP-EMS}s.

\vfill
\end{abstract}
\noindent \textbf{Key words}: Static Passive \emph{EM} Skins; Smart
Electromagnetic Environment; Next-Generation Communications; Multi-Polarization
Skins.

\newpage
\section{Introduction and Rationale\label{sec:Introduction}}

\noindent In the recent years, static passive electromagnetic (\emph{EM})
skins (\emph{SP-EMS}s) have gathered a considerable attention as a
feasible technology to implement Smart Electromagnetic Environments
(\emph{SEME}s) \cite{Yang 2022}-\cite{Barbuto 2022} with minimum
fabrication, installation, and integration costs \cite{Vaquero 2024}-\cite{Liu 2024}.
The macro-scale anomalous wave reflection control, enabled by the
micro-scale structure of \emph{SP-EMS}s \cite{Yang 2019}\cite{Alu 2024},
has been widely exploited in outdoor \cite{Oliveri 2021c}\cite{Oliveri 2023c},
indoor \cite{Vaquero 2024}\cite{Benoni 2023}, and outdoor-to-indoor
\cite{Oliveri 2024} scenarios to improve the coverage as well as
to enhance the quality-of-service (\emph{QoS}) in wireless networks.
Moreover, \emph{SP-EMS}s have been used in \cite{An 2024}\cite{Liu 2024}
as an inexpensive option to validate, in a fixed time snapshot, the
performance of their reconfigurable passive counterparts (\emph{RP-EMS}s),
which are labeled as reconfigurable intelligent surfaces (\emph{RIS}s),
as well.

\noindent Dealing with \emph{SP-EMS}s, the standard \emph{SEME} scenario
generally features a single primary source, which models either a
base station (outdoor case) or an access point (indoor case), that
illuminates the \emph{EMS} aperture \cite{Vaquero 2024}\cite{Benoni 2023}-\cite{Liu 2024}.
Therefore, state-of-the-art design techniques have been usually conceived
to support a single wave-manipulation functionality (e.g., anomalous
focusing or power shaping of the reflected field) \cite{Vaquero 2024}-\cite{Liu 2024}.
Nevertheless, any \emph{EMS} may simultaneously support independent
wave transformations for different illuminations \cite{Yang 2019}\cite{Achouri 2015}
according to the surface electromagnetics theory. This opportunity,
not yet deeply explored in \emph{SEME} scenarios to the best of the
authors' knowledge, can be of great interest in practical applications
where a single \emph{EMS} aperture, featuring multiple functionalities,
could enable multiple service-providers to employ exactly the same
wireless channel, while yielding independent coverage (e.g., Fig.
1).

\noindent Following this line of reasoning, this research work is
devoted to assess the feasibility of multi-functional \emph{SP-EMS}s
for \emph{SEME}s. By leveraging the \emph{polarization diversity}
of the incident waves radiated by two different illumination sources,
but operating at the same frequency (Fig. 1), a method for the synthesis
of a single \emph{EMS} panel \emph{}that supports two independent
wave-manipulation functionalities is introduced and applied to design
multi-polarization (\emph{MP}) \emph{SP-EMS} (\emph{MP-SP-EMS}s).
Towards this end, a simple reference meta-atom is designed first to
enable an accurate and independent control of each polarization component
of the local reflection tensor. Successively, the macro-scale \emph{}synthesis
of multi-polarization (\emph{MP}) \emph{SP-EMS}s (\emph{MP-SP-EMS}s)
is carried out by solving a global optimization problem where a cost
function, which mathematically codes separate requirements for each
polarization, is minimized. More specifically, a customized version
of the system-by-design (\emph{SbD}) technique is implemented to efficiently
handle the huge computational burden of the large-scale problem at
hand.

\noindent To the best of the authors' knowledge, the main innovative
contributions of this paper include (\emph{i}) the design of a low-complexity
unit cell (namely, an \emph{EMS} meta-atom) that supports an independent
control on each $\psi$-th ($\psi\in\left\{ TE,TM\right\} $) component
of the reflection tensor, (\emph{ii}) an \emph{EMS} design approach
to handle two simultaneous functionalities at the same frequency by
exploiting the polarization diversity, and (\emph{iii}) the full-wave
numerical and experimental validation of the concept of multi-polarization
(\emph{MP}) \emph{SP-EMS}s (\emph{MP-SP-EMS}s) in different scenarios
and working conditions.

\noindent The outline of the paper is as follows. After the mathematical
formulation of the synthesis problem at hand, Section \ref{sec:Problem-Formulation}
presents the proposed method for the design of \emph{MP-SP-EMS}s starting
from a simple \emph{EMS} meta-atom layout detailed in Sect. \ref{sub:Meta-Atom-Design}.
Representative results from a set of numerical and experimental experiments
are reported in Sect. \ref{sub:skin full-wave} to assess the feasibility
of a multi-function \emph{EMS} based on polarization diversity as
well as the effectiveness and the robustness of the proposed method
for the synthesis of \emph{MP-SP-EMS}s. Conclusions and final remarks
follow (Sect. \ref{sec:Conclusions-and-Remarks}).

\section{\noindent Problem Formulation and Design Process\label{sec:Problem-Formulation} }

\noindent Let us consider the multi-illumination \emph{SEME} scenario
in Fig. 1(\emph{a}) where two primary sources (e.g., base stations)
illuminate a \emph{MP-SP-EMS} located on the $\left(x,y\right)$-plane
with an incident field $\mathbf{E}^{inc}\left(\mathbf{r}\right)$,
$\mathbf{r}$ ($\mathbf{r}\triangleq\sum_{c=\left\{ x,y,z\right\} }c\widehat{\mathbf{c}}$)
being the position vector. Each source radiates a field modeled as
a plane wave with a different polarization state $\psi$ ($\psi\in\left\{ TE,TM\right\} $),
$\mathbf{E}_{\psi}^{inc}\left(\mathbf{r}\right)$ being the associated
electric field, so that\begin{equation}
\mathbf{E}^{inc}\left(\mathbf{r}\right)=\mathbf{E}_{TE}^{inc}\left(\mathbf{r}\right)+\mathbf{E}_{TM}^{inc}\left(\mathbf{r}\right)=\left(E_{TE}^{inc}\widehat{\mathbf{e}}_{TE}\right)\exp\left(-j\mathbf{k}_{TE}^{inc}\cdot\mathbf{r}\right)+\left(E_{TM}^{inc}\widehat{\mathbf{e}}_{TM}\right)\exp\left(-j\mathbf{k}_{TM}^{inc}\cdot\mathbf{r}\right)\label{eq:incidenti field}\end{equation}
where

\noindent \begin{equation}
\mathbf{k}_{\psi}^{inc}=-k_{0}\left[\sin\left(\theta_{\psi}^{inc}\right)\cos\left(\varphi_{\psi}^{inc}\right)\widehat{\mathbf{x}}+\sin\left(\theta_{\psi}^{inc}\right)\sin\left(\varphi_{\psi}^{inc}\right)\widehat{\mathbf{y}}+\cos\left(\theta_{\psi}^{inc}\right)\widehat{\mathbf{z}}\right]\label{eq:wave-vector}\end{equation}
is the wave vector of the $\psi$-th ($\psi\in\left\{ TE,TM\right\} $)
incident wave component. In (\ref{eq:incidenti field})-(\ref{eq:wave-vector}),
$\widehat{\mathbf{e}}_{\psi}$ is the $\psi$-th ($\psi\in\left\{ TE,TM\right\} $)
mode unit vector, $E_{\psi}^{inc}$ is its complex-valued coefficient,
and $k_{0}$ is the free-space wavenumber. 

\noindent Owing to the linearity of the problem and according to the
surface \emph{EM} theory \cite{Oliveri 2021c}\cite{Yang 2019}\cite{Oliveri 2023b}-\cite{Osipov 2017},
the $\psi$-th ($\psi\in\left\{ TE,TM\right\} $) polarization component
of the field reflected in the far-field region by the \emph{MP-SP-EMS},
$\mathbf{E}_{\psi}^{refl}\left(\mathbf{r}|\mathcal{D}\right)$, is
given by $\mathbf{E}_{\psi}^{refl}\left(\mathbf{r}|\mathcal{D}\right)\triangleq\mathbf{E}_{\psi}^{refl}\left(\theta,\varphi|\mathcal{D}\right)\times\frac{\exp\left(-jk_{0}r\right)}{r}$
where\begin{equation}
\mathbf{E}_{\psi}^{refl}\left(\theta,\varphi|\mathcal{D}\right)=\frac{jk_{0}}{4\pi}\int_{\Omega}\mathbf{J}_{\psi}\left(\mathbf{r}'|\mathcal{D}\right)\exp\left(jk_{0}\widehat{\mathbf{r}}\cdot\mathbf{r}'\right)\mathrm{d}\mathbf{r}',\label{eq:far field}\end{equation}
$\mathbf{J}_{\psi}\left(\mathbf{r}|\mathcal{D}\right)$ being the
corresponding $\psi$-th ($\psi\in\left\{ TE,TM\right\} $) equivalent
surface current induced on the \emph{EMS} aperture $\Omega$ (i.e.,
$\mathbf{r}\in\Omega$) under the $\psi$-th ($\psi\in\left\{ TE,TM\right\} $)
illumination, which is defined as follows\begin{equation}
\mathbf{J}_{\psi}\left(\mathbf{r}|\mathcal{D}\right)\triangleq\widehat{\mathbf{z}}\times\left[\zeta_{0}\widehat{\mathbf{z}}\times\mathbf{J}_{\psi}^{e}\left(\mathbf{r}|\mathcal{D}\right)+\mathbf{J}_{\psi}^{m}\left(\mathbf{r}|\mathcal{D}\right)\right].\label{eq:total currents}\end{equation}
Moreover, $\widehat{\mathbf{r}}=\sin\theta\cos\varphi\widehat{\mathbf{x}}+\sin\theta\sin\varphi\widehat{\mathbf{y}}+\cos\varphi\widehat{\mathbf{z}}$,
$\zeta_{0}$ is the free-space impedance, and $\mathcal{D}$ is the
set of \emph{MP-SP-EMS} \emph{descriptors} that univocally describes
the geometrical and the physical characteristics of the $P\times Q$
meta-atoms of the \emph{EMS}\begin{equation}
\mathcal{D}\triangleq\left\{ \underline{d}_{pq};p=1,...,P;\, q=1,...,Q\right\} ,\label{eq:}\end{equation}
whose ($p$, $q$)-th entry ($p=1,...,P$; $q=1,...,Q$), $\underline{d}_{pq}$
, features $L$ descriptors (i.e., $\underline{d}_{pq}\triangleq\left\{ d_{pq}^{\left(l\right)};\, l=1,...,L\right\} $).

\noindent By exploiting the Love's equivalence principle \cite{Oliveri 2021c}\cite{Yang 2019},
the following relation between the $o$-th ($o\in\left\{ e,m\right\} $)
term of the $\psi$-th ($\psi\in\left\{ TE,TM\right\} $) polarization
component of the induced current, $\mathbf{J}_{\psi}^{o}\left(\mathbf{r}|\mathcal{D}\right)$,
and the \emph{EMS} descriptors, $\mathcal{D}$, is deduced

\noindent \begin{equation}
\left\{ \begin{array}{l}
\mathbf{J}_{\psi}^{e}\left(\mathbf{r}|\mathcal{D}\right)=\frac{1}{\zeta_{0}}\widehat{\mathbf{z}}\times\mathbf{k}_{\psi}^{inc}\times\left[\Gamma_{\psi}\left(\mathbf{r}|\mathcal{D},\mathbf{k}_{\psi}^{inc}\right)E_{\psi}^{inc}\widehat{\mathbf{e}}_{\psi}\right]\\
\mathbf{J}_{\psi}^{m}\left(\mathbf{r}|\mathcal{D}\right)=-\widehat{\mathbf{z}}\times\left[\Gamma_{\psi}\left(\mathbf{r}|\mathcal{D},\mathbf{k}_{\psi}^{inc}\right)E_{\psi}^{inc}\widehat{\mathbf{e}}_{\psi}\right]\end{array}\right.\label{eq:correnti}\end{equation}
where $\Gamma_{\psi}\left(\mathbf{r}|\mathcal{D},\mathbf{k}_{\psi}^{inc}\right)$
is the $\psi$-th ($\psi\in\left\{ TE,TM\right\} $) component of
the local complex reflection tensor of the \emph{MP-SP-EMS} layout
defined by $\mathcal{D}$. Under the local periodicity condition \cite{Yang 2019},
this latter can be approximated as follows\begin{equation}
\Gamma_{\psi}\left(\mathbf{r}|\mathcal{D},\mathbf{k}_{\psi}^{inc}\right)\approx\sum_{p=1}^{P}\sum_{q=1}^{Q}\Gamma_{\psi}\left(\underline{d}_{pq},\mathbf{k}_{\psi}^{inc}\right)\Pi^{pq}\left(\mathbf{r}\right),\label{eq:reflection coefficients}\end{equation}
($\psi\in\left\{ TE,TM\right\} $) $\Pi_{pq}\left(\mathbf{r}\right)$
being the ($p$, $q$)-th ($p=1,...,P$; $q=1,...,Q$) pixel basis
function centered in the barycenter of $pq$-th meta-atom, $\mathbf{r}_{pq}$
($\mathbf{r}_{pq}\in\Omega$).

\noindent It is worth remarking that the relation $\left(\underline{d},\mathbf{k}_{\psi}^{inc}\right)\leftrightarrow\Gamma_{\psi}\left(\underline{d},\mathbf{k}_{\psi}^{inc}\right)$
($\psi\in\left\{ TE,TM\right\} $) can be evaluated, depending on
the layout of the reference meta-atom at hand (i.e., the $L$ descriptors
of $\underline{d}\triangleq\left\{ d^{\left(l\right)};\, l=1,...,L\right\} $),
by using either analytical, or numerical, or hybrid methods as well
as artificial intelligence-based techniques \cite{Yang 2019}\cite{Oliveri 2022b}\cite{Salucci 2018c}.
In this work, a customized learning-by-example (\emph{LBE}) \emph{AI}-based
strategy will be adopted as detailed afterwards. Moreover, the interested
readers should notice that the previous mathematical formulation extends
the framework of \cite{Oliveri 2021c}\cite{Yang 2019} to handle
different impinging directions $\left(\theta_{\psi}^{inc},\varphi_{\psi}^{inc}\right)$
for each $\psi$-th ($\psi\in\left\{ TE,TM\right\} $) incident polarization.

\noindent The problem of synthesizing a \emph{MP-SP-EMS} that reflects
two separate anomalous collimated beams under two {[}one for each
$\psi$-th ($\psi\in\left\{ TE,TM\right\} $) polarization{]} different
illuminations is then stated as follows

\begin{quotation}
\noindent \textbf{\emph{MP-SP-EMS Design Problem}} - Given the incident
field distributions, \{$\mathbf{E}_{\psi}^{inc}\left(\mathbf{r}\right)$;
$\psi\in\left\{ TE,TM\right\} $\}, radiated on the \emph{EMS} aperture
(i.e., $\mathbf{r}\in\Omega$), by the $\psi$-th ($\psi\in\left\{ TE,TM\right\} $)
polarized primary sources, the corresponding reflection directions,
\{$\left(\theta_{\psi}^{refl},\varphi_{\psi}^{refl}\right)$; $\psi\in\left\{ TE,TM\right\} $\},
and a reference layout of the \emph{MP} meta-atom, $\underline{d}$,
characterized by a reflection tensor with polarization components
\{$\Gamma_{\psi}\left(\underline{d},\mathbf{k}_{\psi}^{inc}\right)$;
$\psi\in\left\{ TE,TM\right\} $\}, find the optimal \emph{MP-SP-EMS}
layout (i.e., the set of descriptors, namely the degrees-of-freedom
(\emph{DoF}s) of the synthesis problem at hand, $\mathcal{D}^{opt}$)
that minimizes the following cost function\begin{equation}
\Phi\left(\mathcal{D}\right)=\sum_{\psi\in\left\{ TE,TM\right\} }\alpha_{\psi}\times\left\{ \left|\mathbf{E}_{\psi}^{refl}\left(\theta_{\psi}^{refl},\varphi_{\psi}^{refl}|\mathcal{D}\right)\right|^{2}\right\} ^{-1}\label{eq:anomalous reflection cost function}\end{equation}
(i.e., $\mathcal{D}^{opt}=\arg\left\{ \min_{\mathcal{D}}\left[\Phi\left(\mathcal{D}\right)\right]\right\} $),
$\alpha_{\psi}$ being the $\psi$-th ($\psi\in\left\{ TE,TM\right\} $)
user-defined weighting parameter, while $\left|\mathbf{E}_{\psi}^{refl}\right|\triangleq\sqrt{\sum_{c=\left\{ x,y,z\right\} }\Re^{2}\left\{ \mathbf{E}_{\psi}^{refl}\cdot\widehat{\mathbf{c}}\right\} +j\Im^{2}\left\{ \mathbf{E}_{\psi}^{refl}\cdot\widehat{\mathbf{c}}\right\} }$.
\end{quotation}
\noindent The key challenge of such an \emph{EMS} design problem is
that, unlike the standard synthesis of single anomalous reflection
\emph{EMS}s \cite{Yang 2019}, here no analytical solvers, leveraging
the Generalized Snell's Laws \cite{Yu 2011} and exploiting phase-conjugation,
can be applied owing to the objective of simultaneously maximizing
the reflection of two separate incident waves into independent directions.
As a consequence, the trade-off \emph{EMS} layout, $\mathcal{D}^{opt}$,
turns out to be the solution of a large-scale optimization problem
whose \emph{DoF}s (i.e., the $P\times Q\times L$ entries of $\mathcal{D}^{opt}$
$\triangleq$ \{$\left[d_{pq}^{\left(l\right)}\right]^{opt}$; $l=1,...,L$;
$p=1,...,P$; $q=1,...,Q$\}) must be adjusted to minimize the non-linear
cost function (\ref{eq:anomalous reflection cost function}) accounting
for the \emph{MP} working modality of the \emph{EMS}.

\noindent Towards this end, a customized version of the iterative
\emph{SbD} strategy, discussed in \cite{Oliveri 2021c}\cite{Oliveri 2022b},
is introduced to properly address the \emph{MP-SP-EMS Design Problem}.
More specifically, the process follows the flowchart in Fig. 2 composed
of the following blocks:

\begin{itemize}
\item \noindent \textbf{Solution Space Exploration} (\emph{SSE}) - This
block is aimed at optimizing the \emph{MP-SP-EMS} descriptors by defining
a succession of $S$ iterations ($s$ being the \emph{SbD} iteration
index, $s=1,...,S$) where a set of $G$ trial solutions, \{$\mathcal{D}_{g}^{\left(s\right)}$;
$g=1,...,G$\} ($\mathcal{D}_{g}^{\left(s\right)}$ $\triangleq$
\{$\left.\underline{d}_{pq}\right|_{g}^{\left(s\right)}$; $p=1,...,P$;
$q=1,...,Q$\} being the $g$-th ($g=1,...,G$) solution at the $s$-th
($s=1,...,S$) iteration), evolves towards the global minimum, $\mathcal{D}^{opt}$,
of the objective function (\ref{eq:anomalous reflection cost function}).
Because of the non-linearity of $\Phi\left(\mathcal{D}\right)$, a
global search mechanism based on the \emph{Particle Swarm} paradigm
\cite{Rocca 2009w} is used to update the set/swarm of $G$ trial
solutions at each $s$-th ($s=1,...,S$) iteration;
\item \noindent \textbf{Surface Reflection Coefficient Digital Twin} (\emph{SRCDT})
- This block is devoted to infer the relation $\left(\underline{d},\mathbf{k}_{\psi}^{inc}\right)\leftrightarrow\Gamma_{\psi}\left(\underline{d},\mathbf{k}_{\psi}^{inc}\right)$
for computing $\Gamma_{\psi}\left(\left.\underline{d}_{pq}\right|_{g}^{\left(s\right)},\mathbf{k}_{\psi}^{inc}\right)$
($\psi\in\left\{ TE,TM\right\} $) at every $s$-th ($s=1,...,S$)
\emph{SbD} iteration for each $g$-th ($g=1,...,G$) guess \emph{MP}
meta-atom in the ($p$, $q$)-th ($p=1,...,P$; $q=1,...,Q$) unit
cell of the \emph{EMS} aperture. Since the full-wave computation of
the values of the $\psi$-th ($\psi\in\left\{ TE,TM\right\} $) entries
of the reflection tensor of the $P\times Q\times G\times S$ \emph{MP}
meta-atoms (i.e., \{$\Gamma_{\psi}\left(\left.\underline{d}_{pq}\right|_{g}^{\left(s\right)},\mathbf{k}_{\psi}^{inc}\right)$;
$p=1,...,P$; $q=1,...,Q$; $g=1,...,G$; $s=1,...,S$\}) generated
within the \emph{SbD} iterations would be computationally unfeasible,
each $\psi$-th ($\psi\in\left\{ TE,TM\right\} $) reflection function,
$\Gamma_{\psi}\left(\underline{d},\mathbf{k}_{\psi}^{inc}\right)$,
is approximated with a surrogate relation $\widetilde{\Gamma}_{\psi}\left(\underline{d},\mathbf{k}_{\psi}^{inc}\right)$
predicted by a \emph{Digital Twin} (\emph{DT})%
\footnote{\noindent It is worth noticing that in this case, unlike \cite{Oliveri 2021c}\cite{Oliveri 2022b}
where the generalized sheet transition condition technique has been
adopted, the \emph{DT} predicts the local reflection coefficients
as a function of both the unit-cell geometry, $\underline{d}$, and
the incidence-wave directions, \{$\mathbf{k}_{\psi}^{inc}$; $\psi\in\left\{ TE,TM\right\} $\},
which here are different ($\mathbf{k}_{TE}^{inc}\ne\mathbf{k}_{TM}^{inc}$).%
} based on the statistical learning \emph{Ordinary Kriging} (\emph{OK})
method \cite{Oliveri 2021c}\cite{Oliveri 2022b}\cite{Salucci 2018c}\cite{Oliveri 2022c}
and trained with a set of \emph{N} known pairs/\emph{examples}, \{$\left[\underline{d},\mathbf{k}_{\psi}^{inc}\right]_{n}$,
$\Gamma_{\psi}\left(\left[\underline{d},\mathbf{k}_{\psi}^{inc}\right]_{n}\right)$
($\psi\in\left\{ TE,TM\right\} $); $n=1,...,N$\}; 
\item \noindent \textbf{Surface-Current Evaluation} (\emph{SCE}) - For each
$g$-th ($g=1,...,G$) trial \emph{EMS} layout at the $s$-th ($s=1,...,S$)
\emph{SbD} iteration, $\mathcal{D}_{g}^{\left(s\right)}$, this block
implements (\ref{eq:correnti}) to determine the $o$-th ($o\in\left\{ e,m\right\} $)
term of the $\psi$-th ($\psi\in\left\{ TE,TM\right\} $) polarization
component of the induced current, $\mathbf{J}_{\psi}^{o}\left(\mathbf{r}|\mathcal{D}\right)$,
starting from the knowledge of the $\psi$-th ($\psi\in\left\{ TE,TM\right\} $)
incident field, $\mathbf{E}_{\psi}^{inc}\left(\mathbf{r}\right)$,
and the corresponding reflection coefficient, $\Gamma_{\psi}\left(\mathbf{r}|\mathcal{D}_{g}^{\left(s\right)},\mathbf{k}_{\psi}^{inc}\right)$,
through (\ref{eq:reflection coefficients}) by inputting the $P\times Q$
values of \{$\Gamma_{\psi}\left(\left.\underline{d}_{pq}\right|_{g}^{\left(s\right)},\mathbf{k}_{\psi}^{inc}\right)$;
$p=1,...,P$; $q=1,...,Q$\} predicted in the \emph{SRCDT} block;
\item \noindent \textbf{Far Field Computation} (\emph{FFC}) - This block
computes $\mathbf{E}_{\psi}^{refl}\left(\theta,\varphi|\mathcal{D}_{g}^{\left(s\right)}\right)$
($\psi\in\left\{ TE,TM\right\} $) by first substituting in (\ref{eq:total currents})
the $\psi$-th ($\psi\in\left\{ TE,TM\right\} $) currents, \{$\mathbf{J}_{\psi}^{o}\left(\mathbf{r}|\mathcal{D}_{g}^{\left(s\right)}\right)$;
$o\in\left\{ e,m\right\} $\} determined in the \emph{SCE} block and
finally inputting the result in (\ref{eq:far field});
\item \noindent \textbf{Cost Function Evaluation} (\emph{CFE}) - This block
implements (\ref{eq:anomalous reflection cost function}) to compute
the cost function value of the $g$-th ($g=1,...,G$) trial \emph{EMS}
layout guessed at the $s$-th ($s=1,...,S$) \emph{SbD} iteration,
$\Phi\left(\mathcal{D}_{g}^{\left(s\right)}\right)$, from the value
of $\mathbf{E}_{\psi}^{refl}\left(\theta,\varphi|\mathcal{D}_{g}^{\left(s\right)}\right)$
($\psi\in\left\{ TE,TM\right\} $) computed in the \emph{FFC} block.
\end{itemize}

\section{\noindent Numerical and Experimental Validation\label{sec:Results}}

\noindent The objective of this section is twofold. In Sect. \ref{sub:Meta-Atom-Design},
the design of a low-complexity \emph{MP} meta-atom suitable for the
implementation of a \emph{MP-SP-EMS} structure is assessed. Otherwise,
Section \ref{sub:skin full-wave} is devoted to analyze the wave manipulation
performance of a set of multi-functional \emph{EMS} layouts yielded
by properly arranging, on the \emph{EMS} aperture and according to
the synthesis method detailed in Sect. \ref{sec:Problem-Formulation},
suitable parametric-variations of the reference meta-atom of Sect.
\ref{sub:Meta-Atom-Design}. Towards this end, the analytic predictions
of the behavior of the synthesized \emph{EMS}s are compared with full-wave
numerical simulations performed in \emph{Ansys HFSS} \cite{HFSS 2021}
and experimental measurements (Sect. \ref{sub:skin full-wave}).

\noindent In the following, the primary sources have been assumed
to operate at $f_{0}=28$ {[}GHz{]} and to illuminate the \emph{EMS}
panel with unitary fields (i.e., $E_{\psi}^{inc}=1$ {[}V/m{]}, $\psi\in\left\{ TE,TM\right\} $)
by setting $\varphi_{\psi}^{inc}=\varphi_{\psi}^{refl}=0$ {[}deg{]}
($\psi\in\left\{ TE,TM\right\} $), as well. As for the manufacturing
of the \emph{EMS} meta-atom and the arising \emph{MP-SP-EMS}s, a $H=5.1\times10^{-4}$
{[}m{]}-thick Rogers \emph{RO3003} sheet with dielectric relative
permittivity and dielectric loss tangent equal to $\varepsilon_{r}=3.0$
and $\tan\delta=1.0\times10^{-3}$, respectively, has been chosen
for the \emph{EMS} substrate, while all the metallizations of the
\emph{EMS} layout have been realized with a standard copper layer
of thickness $35$ {[}$\mu$m{]}.

\subsection{Meta-Atom Design\label{sub:Meta-Atom-Design}}

\noindent To prove the feasibility of the \emph{MP-SP-EMS} concept,
a unit cell geometry (i.e., $\underline{d}$) supporting an \emph{independent}
control on each $\psi$-th ($\psi\in\left\{ TE,TM\right\} $) polarization-component
of the reflection tensor, $\Gamma_{\psi}\left(\underline{d},\mathbf{k}_{\psi}^{inc}\right)$,
with a minimum topological complexity (e.g., $L$ $\downarrow$) is
needed. However, state-of-the-art \emph{EMS} atoms, based on square
metallizations on single-layer substrates, are not suitable because
of their single polarization features \cite{Oliveri 2021c}\cite{Benoni 2023}\cite{Liu 2024}.
On the other hand, advanced patterning layouts as in \cite{Vaquero 2024}
were not adopted since too complex and expensive for a cheap and accessible
prototyping especially in view of (future) wide deployments. Starting
from these considerations, the meta-atom sketched in Fig. 1(\emph{b}),
which consists of a single rectangular patch printed on a single-layer
substrate and is characterized by $L=2$ \emph{DoF}s, has been selected.
To evaluate the achievable magnitude/phase control on each $\psi$-th
($\psi\in\left\{ TE,TM\right\} $) reflected field component, such
an \emph{EMS} unit-cell has been numerically modeled in \emph{Ansys
HFSS} when replicated in each sub-domain of a square lattice with
periodicity $\Delta x=\Delta y=0.4\lambda$ by setting periodic boundary
conditions under the local periodicity approximation.

\noindent Figure 3 gives the $\psi$-th ($\psi\in\left\{ TE,TM\right\} $)
reflection coefficient, $\Gamma_{\psi}\left(\underline{d},\mathbf{k}_{\psi}^{inc}\right)$,
as a function of the $L$ descriptors of the meta-atom (i.e., $d^{\left(1\right)}$
and $d^{\left(2\right)}$ $\to$ $L=2$) when this latter is illuminated
from broadside by the two co-located primary sources {[}i.e., $\theta_{\psi}^{inc}=0$
{[}deg{]} ($\psi\in\left\{ TE,TM\right\} $){]}. Regardless of the
$\psi$-th ($\psi\in\left\{ TE,TM\right\} $) polarization of the
source and the topological variations of the atom {[}i.e., geometrical
variations of each $l$-th ($l=1,...,L$) descriptor, $d^{\left(l\right)}${]},
the reflection efficiency is always greater than $90$ \% ($98$ \%
in a wide part of the range of variation of the $L=2$ variables),
the magnitude of any $\psi$-th ($\psi\in\left\{ TE,TM\right\} $)
reflection coefficient being always greater than $-0.5$ {[}dB{]}
(i.e., $\left|\Gamma_{\psi}\left(\underline{d},\mathbf{k}_{\psi}^{inc}\right)\right|>-0.5$
{[}dB{]}) and generally close to $-0.1$ {[}dB{]} {[}Fig. 3(\emph{a})
and Fig. 3(\emph{c}){]}. Moreover, the value of the phase coverage
index, $\Delta\Gamma_{\psi}\left(\mathbf{k}_{\psi}^{inc}\right)$
($\Delta\Gamma_{\psi}\left(\mathbf{k}_{\psi}^{inc}\right)$ $\triangleq$
$\max_{\underline{d}}\left[\angle\Gamma_{\psi}\left(\underline{d},\mathbf{k}_{\psi}^{inc}\right)\right]$
$-$ $\min_{\underline{d}}\left[\angle\Gamma_{\psi}\left(\underline{d},\mathbf{k}_{\psi}^{inc}\right)\right]$),
for both polarization components ($\psi\in\left\{ TE,TM\right\} $)
turns out to be close to the full $360$ {[}deg{]} range (i.e., $\Delta\Gamma_{\psi}\left(\mathbf{k}_{\psi}^{inc}\right)\approx325$
{[}deg{]}, $\psi\in\left\{ TE,TM\right\} $) since almost all the
phase values of the reflection coefficients between $-180$ {[}deg{]}
and $180$ {[}deg{]} can be yielded by an available/admissible setup
of the atom variables, \{$d^{\left(l\right)}$; $l=1,...,L$\} {[}Fig.
3(\emph{b}) and Fig. 3(\emph{d}){]}. 

\noindent These outcomes prove that the unit-cell model in Fig. 1(\emph{b})
fulfils the conditions for yielding anomalous wave-manipulation properties
with minimum phase-quantization distortions. Finally, it is also worth
mentioning that, as expected from the theoretical viewpoint, both
the magnitude and the phase response of such meta-atom exhibit a polarization
symmetry (i.e., $\left|\Gamma_{TE}\left(d_{1},d_{2};\mathbf{k}_{TE}^{inc}\right)\right|=\left|\Gamma_{TM}\left(d_{2},d_{1};\mathbf{k}_{TE}^{inc}\right)\right|$
{[}Fig. 3(\emph{a}) and Fig. 3(\emph{c}){]} and $\angle\Gamma_{TE}\left(d_{1},d_{2};\mathbf{k}_{TE}^{inc}\right)=\angle\Gamma_{TM}\left(d_{2},d_{1};\mathbf{k}_{TE}^{inc}\right)$
{[}Fig. 3(\emph{b}) and Fig. 3(\emph{d}){]}) owing to the rotational
symmetry of its layout {[}Fig. 1(\emph{b}){]}.

\subsection{\emph{EMS} Design and Validation\label{sub:skin full-wave}}

\noindent Starting from the meta-atom described in Sect. \ref{sub:Meta-Atom-Design},
the design process of \emph{MP-SP-EMS}s has been carried out by setting
$\alpha_{\psi}=1.0$ ($\psi\in\left\{ TE,TM\right\} $) and choosing
the values of the calibration parameters of the \emph{SbD}-based optimization
according to the guidelines in \cite{Oliveri 2021c}, namely $S=10^{4}$
and $G=10^{2}$.

\noindent The synthesis of a square $P\times Q=20\times20$ \emph{EMS}
of side $\mathcal{L}\approx8.57\times10^{-2}$ has been addressed
first (\emph{Test Case} \emph{1}) by co-locating the pair of primary
sources so that $\theta_{\psi}^{inc}=0$ {[}deg{]} ($\psi\in\left\{ TE,TM\right\} $)
and requiring anomalous reflections towards $\theta_{TE}^{refl}=30$
{[}deg{]} and $\theta_{TM}^{refl}=-40$ {[}deg{]}, respectively. The
sketch of the \emph{HFSS} model derived from the descriptors, $\mathcal{D}^{opt}$,
of the synthesized \emph{MP-SP-EMS} is reported in Fig. 4. As expected,
the arising arrangement of the \emph{EMS} meta-atoms is characterized
by a non-uniform distribution of irregular rectangular shapes (Fig.
4). Indeed, it is the trade-off result for yielding two co-existing
target functionalities, each one requiring a different distribution,
on the same \emph{EMS} aperture, of the phase of the local reflection
coefficient (i.e., $\angle\Gamma_{TE}\left(\mathbf{r}|\mathcal{D},\mathbf{k}_{TE}^{inc}\right)$
{[}Fig. 5(\emph{a}){]} and $\angle\Gamma_{TM}\left(\mathbf{r}|\mathcal{D},\mathbf{k}_{TM}^{inc}\right)$
{[}Fig. 5(\emph{b}){]}). The analytically-simulated fields reflected
by such an \emph{EMS} are shown in Figs. 5(\emph{c})-5(\emph{d}) by
using the ($u$, $v$) direction cosines representation ($u\triangleq\sin\theta\cos\varphi$,
$v\triangleq\sin\theta\sin\varphi$). More in detail, each $\psi$-th
($\psi\in\left\{ TE,TM\right\} $) pattern has been computed separately
for the two co-located illuminations by setting either ($E_{TE}^{inc}=1$,
$E_{TM}^{inc}=0$) {[}Fig. 5(\emph{c}){]} or ($E_{TE}^{inc}=0$, $E_{TM}^{inc}=1$)
{[}Fig. 5(\emph{d}){]}. As it can be observed, the \emph{MP-SP-EMS}
implements a well controlled anomalous reflection along $\theta_{TE}^{refl}=30$
{[}deg{]} {[}$\to$ $\left(u,v\right)=\left(0.5,0.0\right)$ - Fig.
5(\emph{c}){]} as well as towards $\theta_{TM}^{refl}=-40$ {[}deg{]}
{[}$\to$ $\left(u,v\right)=\left(-0.64,0.0\right)$ - Fig. 5(\emph{d}){]}
when illuminated by the \emph{TE}- or the \emph{TM-}source, respectively.
Moreover, despite the elementary layout of the meta-atom in Fig. 1(\emph{b}),
the synthesized \emph{EMS} reflects a field pattern with well-controlled
secondary lobes, not only along the $\varphi=0$ cut (Fig. 6), but
in the whole visible range $\Theta$ ($\Theta$ $\triangleq$ \{$\left(u,v\right)$
$\left|u^{2}+v^{2}=1\right.$\}) {[}Figs. 5(\emph{c})-5(\emph{d}){]}.
Furthermore, the faithful matching between analytical predictions
and full-wave \emph{Ansys HFSS} simulations (Fig. 6) demonstrates
the reliability of the local periodicity approximation (see Sect.
\ref{sec:Problem-Formulation}) as well as the marginal impact of
both the (non-uniform) mutual coupling among the \emph{EMS} meta-atoms
and the edge truncation effects of the finite-size \emph{EMS} aperture.
These results assess the feasibility of a multi-functional \emph{EMS}
that offers a separate response for each of the two independent incident
fields thanks to the polarization diversity mechanism.

\noindent In the second experiment of the first test case, the direction
of incidence on the \emph{EMS} panel of the impinging wave has been
varied from broadside {[}$\theta_{\psi}^{inc}=0$ {[}deg{]} ($\psi\in\left\{ TE,TM\right\} $){]}
to $\theta_{\psi}^{inc}=-30$ {[}deg{]} ($\psi\in\left\{ TE,TM\right\} $),
while keeping unaltered the reflection directions to $\theta_{TE}^{refl}=30$
{[}deg{]} and $\theta_{TM}^{refl}=-40$ {[}deg{]}. The $P\times Q$
atoms layout of the synthesized \emph{EMS} is shown in Fig. 7(\emph{a}),
while the plots of both analytically-computed and full-wave simulated
reflected field patterns in the $\varphi=0$ {[}deg{]} plane are given
in Fig. 7(\emph{b}) to assess the flexibility/generality of the proposed
technological solution in achieving the desired multi-functional properties
for different incidence directions {[}i.e., $\theta_{\psi}^{inc}\neq\left(0,0\right)$
{[}deg{]}, $\psi\in\left\{ TE,TM\right\} ${]}.

\noindent The scalability of the proposed design concept to wider
arrangements is evaluated next (\emph{Test Case 2}), but maintaining
the same incidence/reflection conditions of Fig. 7. By considering
two different \emph{EMS} apertures composed of $P\times Q=30\times30$
($\to$ $\mathcal{L}\approx1.28\times10^{-1}$ {[}m{]}) and $P\times Q=40\times40$
($\to$ $\mathcal{L}\approx1.71\times10^{-2}$ {[}m{]}) unit-cells,
the optimized \emph{EMS} layouts in Fig. 8 have been obtained. Regardless
of the \emph{EMS} size, a reliable polarization-based coverage control
with well controlled sidelobes is implemented as highlighted in Fig.
9(\emph{a}) ($P\times Q=30\times30$) and Fig. 9(\emph{b}) ($P\times Q=40\times40$)
where the reflection patterns in the $\varphi=0$ {[}deg{]} cut are
reported.

\noindent Concerning the computational load of the \emph{EMS} synthesis,
even though the \emph{CPU}-time grows linearly with the number of
meta-atoms of the \emph{EMS}, the optimization process detailed in
Sect. \ref{sec:Problem-Formulation} turns out to be very efficient
and competitive to handle standard \emph{EMS} apertures. For instance,
the design of a $P\times Q=20\times20$ elements \emph{EMS} requires%
\footnote{\noindent The simulation time is related to an execution of a non-optimized
MATLAB serial implementation of the synthesis algorithm on a standard
laptop with a single-core $1.6$ GHz CPU.%
} less than $50$ {[}sec{]} (i.e., $\left.\Delta t\right|_{P\times Q=20\times20}\approx4.1\times10^{1}$
{[}sec{]}), while the synthesis of a wider \emph{EMS} of size $P\times Q=40\times40$
needs less than $3$ {[}min{]} (i.e., $\left.\Delta t\right|_{P\times Q=40\times40}\approx1.7\times10^{2}$
{[}sec{]}). 

\noindent The final representative example (\emph{Test Case 3}) considers
separated (i.e., $\theta_{TE}^{inc}\ne\theta_{TM}^{inc}$) primary
sources. More specifically, the $P\times Q=40\times40$ ($\to$ $\mathcal{L}\approx1.71\times10^{-2}$
{[}m{]}) \emph{EMS} in Fig. 10(\emph{a}) has been designed subject
to the following set of constraints: $\theta_{TE}^{inc}=-30$ {[}deg{]},
$\theta_{TM}^{inc}=40$ {[}deg{]}, $\theta_{TE}^{refl}=20$ {[}deg{]},
and $\theta_{TM}^{refl}=-20$ {[}deg{]} {[}$\varphi_{\psi}^{inc}=\varphi_{\psi}^{refl}=0.0$
($\psi\in\left\{ TE,TM\right\} $){]}. Moreover, a prototype of the
synthesized \emph{EMS} layout has been built (Fig. 10) to also experimentally
assess the effectiveness of \emph{MP-SP-EMS}s as well as to evaluate
the impact of the fabrication tolerances on the reflection performance
of the \emph{EMS}. In particular, the arising arrangement of meta-atoms
has been manufactured on a $H=5.1\times10^{-4}$ {[}m{]}-thick Rogers
\emph{RO3003} with a standard \emph{PCB} prototyping methodology {[}Fig.
10(\emph{b}){]}. The measurement process has been performed in a semi-anechoic
chamber by using a \emph{RF} source operating at $f_{0}=28$ {[}GHz{]}
and a \emph{}standard horn antenna with $15$ {[}dB{]} gain has been
used for both the primary source and the \emph{EM} field probe. The
transmitting antenna has been installed on a fixed support located
$\approx100$ {[}cm{]} away from the \emph{MP-SP-EMS}, while both
the \emph{EMS} and the illuminator have been located on a mechanically
rotating platform to automatically collect the reflected pattern samples
in the whole azimuth plane (Fig. 11). The reflected field values of
each $\psi$-th ($\psi\in\left\{ TE,TM\right\} $) polarization have
been measured in different and subsequent sessions by suitably rotating
the horn antennas.

\noindent Figure 11 shows an excellent agreement among analytically-simulated,
numerically-computed, and measured (normalized to the transmitting/receiving
antenna gain) reflection patterns. Such an outcome confirms, also
experimentally, the possibility to implement independent functionalities
with a single \emph{SP-EMS} that serves two (also separated) orthogonally-polarized
primary sources. Moreover, it is worth pointing out that the \emph{EMS}
performance/functionalities turn out marginally affected by the unavoidable
fabrication tolerances (Fig. 11) thanks to the intrinsic robustness
and the simplicity of the \emph{EMS} meta-atom in Fig. 1(\emph{b}).

\section{\noindent Conclusions and Final Remarks\label{sec:Conclusions-and-Remarks}}

\noindent The concept of multi-functional \emph{SP-EMS}s for \emph{SEME}s
has been demonstrated by yielding an independent control of the reflection
of two incident fields at the same frequency with orthogonal \emph{}polarizations.
The design of the micro-scale features of the \emph{EMS}, which is
requested to simultaneously support two macro-scale functionalities
in reflection, has been recast as a global optimization problem then
solved with a customized \emph{SbD}-based approach.

\noindent From the numerical and experimental assessment, the following
main outcomes can be drawn:

\begin{itemize}
\item \noindent it is possible to design a \emph{MP-SP-EMS} that \emph{}enables
an independent control of the anomalous reflection of the waves radiated
towards the \emph{EMS} panel by two differently-polarized sources
working at the same frequency {[}e.g., Figs. 5(\emph{c})-5(\emph{d}),
Fig. 7(\emph{b}), Fig. 9, and Fig. 11{]};
\item the physical implementation of a \emph{MP-SP-EMS} requires neither
complex meta-atom layouts nor expensive (e.g., multi-layers) substrates
as confirmed by the manufactured prototype (Fig. 10);
\item the proposed design process (Sect. \ref{sec:Problem-Formulation})
turns out to be effective (faithfully fulfilling the design constraints),
reliable (the analytic predictions adequately matching both numerical
simulations and experimental measurements), and numerically efficient
(the \emph{CPU}-time being affordable in handling standard - even
large - apertures for \emph{SEME}s).
\end{itemize}
\noindent Future works, beyond the scope of the current manuscript,
will be aimed at (\emph{i}) generalizing the proposed multi-functional
working principle to \emph{RP-EMS}s and (\emph{ii}) exploiting the
available functionalities to enable integrated sensing and communication
(\emph{ISAC}) services.

\section*{\noindent Acknowledgements}

\noindent This work benefited from the networking activities carried
out within the Project ICSC National Centre for HPC, Big Data and
Quantum Computing (CN HPC) funded by the European Union - NextGenerationEU
within the PNRR Program (CUP: E63C22000970007), the Project DICAM-EXC
funded by the Italian Ministry of Education, Universities and Research
(MUR) (Departments of Excellence 2023-2027, grant L232/2016), the
Project INSIDE-NEXT - Indoor Smart Illuminator for Device Energization
and Next-Generation Communications funded by the Italian Ministry
for Universities and Research within the PRIN 2022 Program (CUP: E53D23000990001),
and the Project AURORA - Smart Materials for Ubiquitous Energy Harvesting,
Storage, and Delivery in Next Generation Sustainable Environments
funded by the Italian Ministry for Universities and Research within
the PRIN-PNRR 2022 Program, and the Project Partnership on Telecommunications
of the Future (PE00000001 - program RESTART), funded by the European
Union under the Italian National Recovery and Resilience Plan (NRRP)
of NextGenerationEU (CUP: E63C22002040007). A. Massa wishes to thank
E. Vico and L. Massa for the never-ending inspiration, support, guidance,
and help.

\newpage
\section*{FIGURE CAPTIONS}

\begin{itemize}
\item \textbf{Figure 1.} \emph{Problem Scenario} - Sketch of (\emph{a})
the reference \emph{SEME} and (\emph{b}) the meta-atom geometry.
\item \textbf{Figure 2.} \emph{EMS Design Method} - Flowchart and building
blocks.
\item \textbf{Figure 3.} \emph{Meta-Atom Analysis} (\{$\theta_{\psi}^{inc}=\varphi_{\psi}^{inc}=\left(0,0\right)$
{[}deg{]}; $\psi\in\left\{ TE,TM\right\} $\}, $L=2$) - Behaviour
of (\emph{a})(\emph{c}) the magnitude and (\emph{b})(\emph{d}) the
phase of $\Gamma_{\psi}\left(\underline{d},\mathbf{k}_{\psi}^{inc}\right)$
versus the $L$ descriptors of the meta-atom, \{$d^{\left(l\right)}$;
$l=1,...,L$\}: (\emph{a})(\emph{b}) $\psi\to TE$ and (\emph{c})(\emph{d})
$\psi\to TM$.
\item \textbf{Figure 4.} \emph{Numerical Validation} ($P\times Q=20\times20$,
$\theta_{\psi}^{inc}=\varphi_{\psi}^{inc}=\varphi_{\psi}^{refl}=0$
{[}deg{]} ($\psi\in\left\{ TE,TM\right\} $), $\theta_{TE}^{refl}=30$
{[}deg{]}, $\theta_{TM}^{refl}=-40$ {[}deg{]}) - Sketch of the \emph{HFSS}
model of the synthesized \emph{MP-SP-EMS.}
\item \textbf{Figure 5.} \emph{Numerical Validation} ($P\times Q=20\times20$,
$\theta_{\psi}^{inc}=\varphi_{\psi}^{inc}=\varphi_{\psi}^{refl}=0$
{[}deg{]} ($\psi\in\left\{ TE,TM\right\} $), $\theta_{TE}^{refl}=30$
{[}deg{]}, $\theta_{TM}^{refl}=-40$ {[}deg{]}) - Plots of (\emph{a})(\emph{b})
the phase of the $\psi$-th component of the local reflection coefficient,
$\angle\Gamma_{\psi}\left(\mathbf{r}|\mathcal{D},\mathbf{k}_{\psi}^{inc}\right)$,
in the \emph{EMS} aperture $\Omega$, (\emph{c})(\emph{d}) the magnitude
of the $\psi$-th component of reflected field pattern, $\left|\mathbf{E}_{\psi}^{refl}\left(u,v\right)\right|$,
in the visible range $\Theta$ ($\Theta$ $\triangleq$ \{$\left(u,v\right)$
$\left|u^{2}+v^{2}=1\right.$\}): (\emph{a})(\emph{c}) $\psi\to TE$
and (\emph{b})(\emph{d}) $\psi\to TM$.
\item \textbf{Figure 6.} \emph{Numerical Validation} ($P\times Q=20\times20$,
$\theta_{\psi}^{inc}=\varphi_{\psi}^{inc}=\varphi_{\psi}^{refl}=0$
{[}deg{]} ($\psi\in\left\{ TE,TM\right\} $), $\theta_{TE}^{refl}=30$
{[}deg{]}, $\theta_{TM}^{refl}=-40$ {[}deg{]}) - Plot of the magnitude
of the \{$\psi$-th ($\psi\in\left\{ TE,TM\right\} $\} components
of the reflected field pattern, \{$\left|\mathbf{E}_{\psi}^{refl}\left(u,v\right)\right|$
($\psi\in\left\{ TE,TM\right\} $)\}, in the $\varphi=0$ {[}deg{]}
plane.
\item \textbf{Figure 7.} \emph{Numerical Validation} ($P\times Q=20\times20$,
\{$\theta_{\psi}^{inc}=-30$ {[}deg{]}, $\varphi_{\psi}^{inc}=\varphi_{\psi}^{refl}=0$
{[}deg{]}; $\psi\in\left\{ TE,TM\right\} $\}, $\theta_{TE}^{refl}=30$
{[}deg{]}, $\theta_{TM}^{refl}=-40$ {[}deg{]}) - Pictures of (\emph{a})
the sketch of the \emph{HFSS} model of the synthesized \emph{MP-SP-EMS}
and (\emph{b}) the magnitude of the \{$\psi$-th ($\psi\in\left\{ TE,TM\right\} $\}
components of the reflected field pattern, \{$\left|\mathbf{E}_{\psi}^{refl}\left(u,v\right)\right|$
($\psi\in\left\{ TE,TM\right\} $)\}, in the $\varphi=0$ {[}deg{]}
plane.
\item \textbf{Figure 8.} \emph{Numerical Validation} (\{$\theta_{\psi}^{inc}=-30$
{[}deg{]}, $\varphi_{\psi}^{inc}=\varphi_{\psi}^{refl}=0$ {[}deg{]};
$\psi\in\left\{ TE,TM\right\} $\}, $\theta_{TE}^{refl}=30$ {[}deg{]},
$\theta_{TM}^{refl}=-40$ {[}deg{]}) \emph{-} Sketches of the \emph{HFSS}
model of the synthesized \emph{MP-SP-EMS} with (\emph{a}) $P\times Q=30\times30$
and (\emph{a}) $P\times Q=40\times40$ meta-atoms.
\item \textbf{Figure 9.} \emph{Numerical Validation} (\{$\theta_{\psi}^{inc}=-30$
{[}deg{]}, $\varphi_{\psi}^{inc}=\varphi_{\psi}^{refl}=0$ {[}deg{]};
$\psi\in\left\{ TE,TM\right\} $\}, $\theta_{TE}^{refl}=30$ {[}deg{]},
$\theta_{TM}^{refl}=-40$ {[}deg{]}) \emph{-} Plots of the magnitude
of the \{$\psi$-th ($\psi\in\left\{ TE,TM\right\} $\} components
of the reflected field pattern, \{$\left|\mathbf{E}_{\psi}^{refl}\left(u,v\right)\right|$
($\psi\in\left\{ TE,TM\right\} $)\}, in the $\varphi=0$ {[}deg{]}
plane in correspondence with the synthesized \emph{MP-SP-EMS} with
(\emph{a}) $P\times Q=30\times30$ {[}Fig. 8(\emph{a}){]} and (\emph{a})
$P\times Q=40\times40$ {[}Fig. 8(\emph{b}){]} meta-atoms.
\item \textbf{Figure 10.} \emph{Experimental Validation} ($P\times Q=40\times40$,
\emph{}$\theta_{TE}^{inc}=-30$ {[}deg{]}, $\theta_{TM}^{inc}=40$
{[}deg{]}, $\theta_{TE}^{refl}=20$ {[}deg{]}, $\theta_{TM}^{refl}=-20$
{[}deg{]}, \{$\varphi_{\psi}^{inc}=\varphi_{\psi}^{refl}=0$ {[}deg{]};
$\psi\in\left\{ TE,TM\right\} $\}) - Pictures of (\emph{a}) the \emph{EMS}
prototype and (\emph{b}) the detail of the \emph{EMS} metallizations.
\item \textbf{Figure 11.} \emph{Experimental Validation} ($P\times Q=40\times40$,
\emph{}$\theta_{TE}^{inc}=-30$ {[}deg{]}, $\theta_{TM}^{inc}=40$
{[}deg{]}, $\theta_{TE}^{refl}=20$ {[}deg{]}, $\theta_{TM}^{refl}=-20$
{[}deg{]}, \{$\varphi_{\psi}^{inc}=\varphi_{\psi}^{refl}=0$ {[}deg{]};
$\psi\in\left\{ TE,TM\right\} $\}) - Plot of the magnitude of the
$\psi$-th component of reflected field pattern, $\left|\mathbf{E}_{\psi}^{refl}\left(u,v\right)\right|$,
in the $\varphi=0$ {[}deg{]} plane: (\emph{a}) $\psi\to TE$ and
(\emph{b}) $\psi\to TM$.
\end{itemize}
\noindent ~

\newpage
\begin{center}~\vfill\end{center}

\begin{center}\begin{tabular}{c}
\includegraphics[%
  width=0.90\columnwidth,
  keepaspectratio]{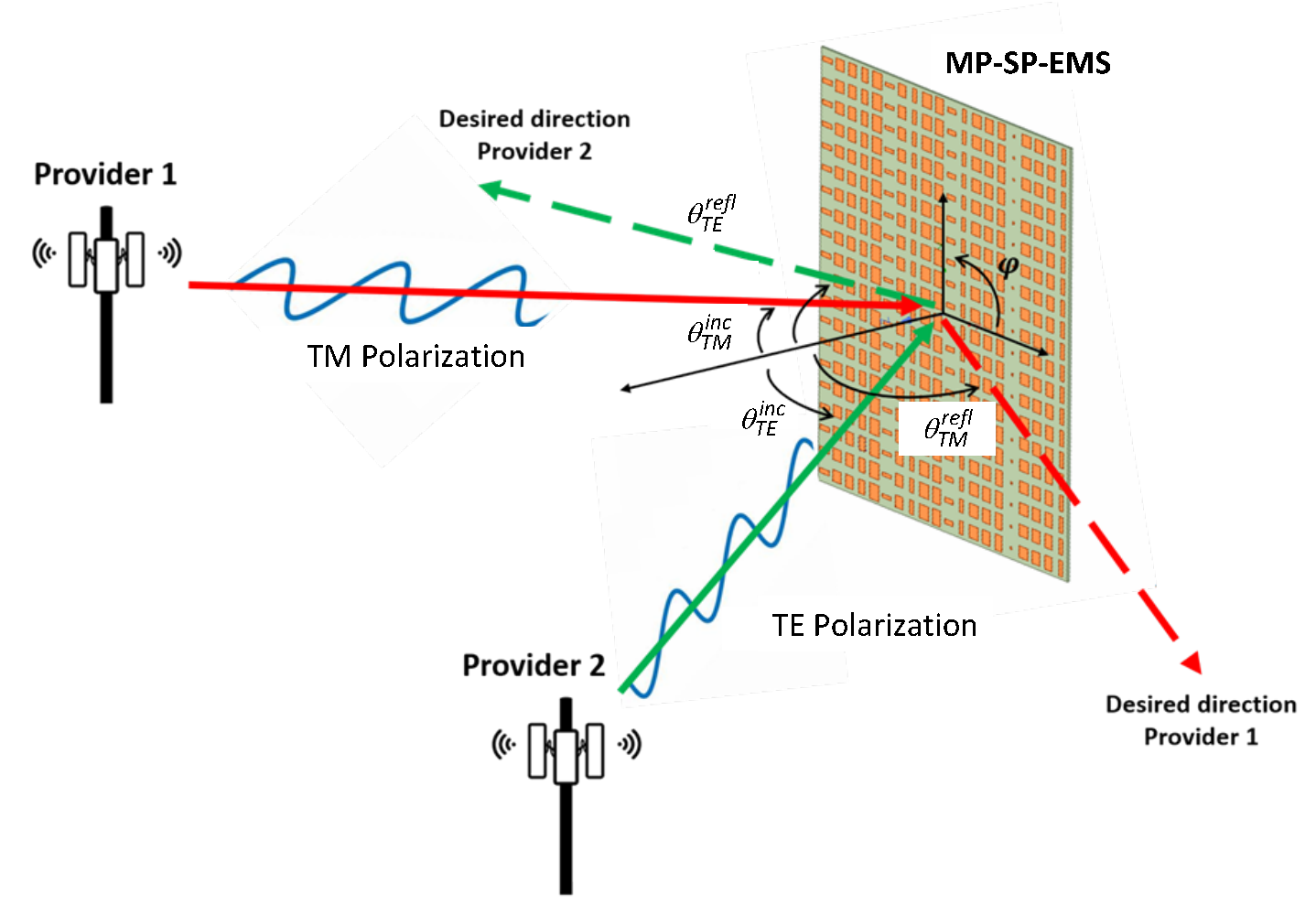}\tabularnewline
(\emph{a})\tabularnewline
\includegraphics[%
  width=0.90\columnwidth,
  keepaspectratio]{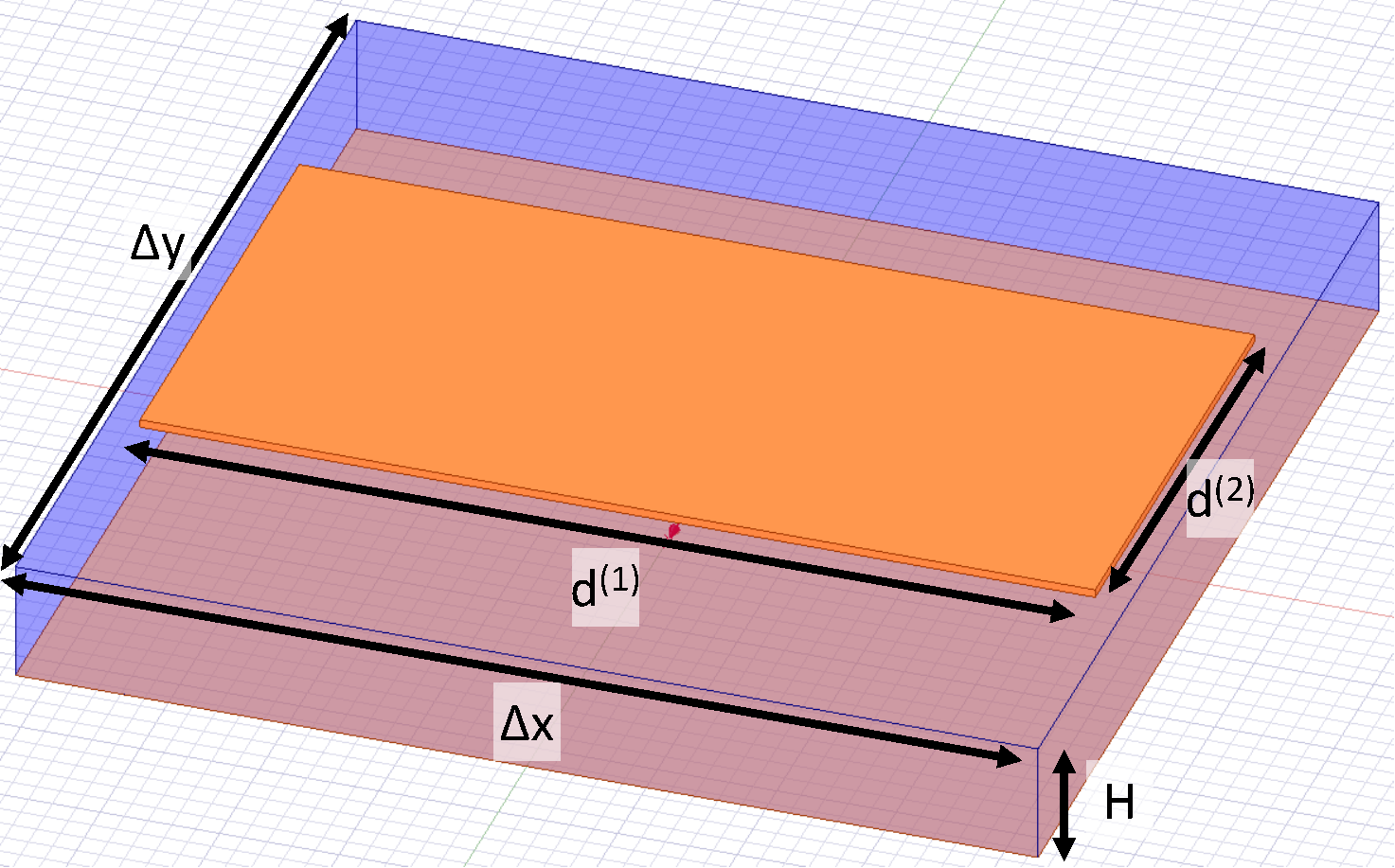}\tabularnewline
(\emph{b})\tabularnewline
\end{tabular}\end{center}

\begin{center}~\vfill\end{center}

\begin{center}\textbf{Fig. 1 - G. Oliveri} \textbf{\emph{et al.,}}
{}``Multi-Functional Polarization-Based Coverage Control through
...''\end{center}

\newpage
\begin{center}~\vfill\end{center}

\begin{center}\includegraphics[%
  width=0.90\columnwidth,
  keepaspectratio]{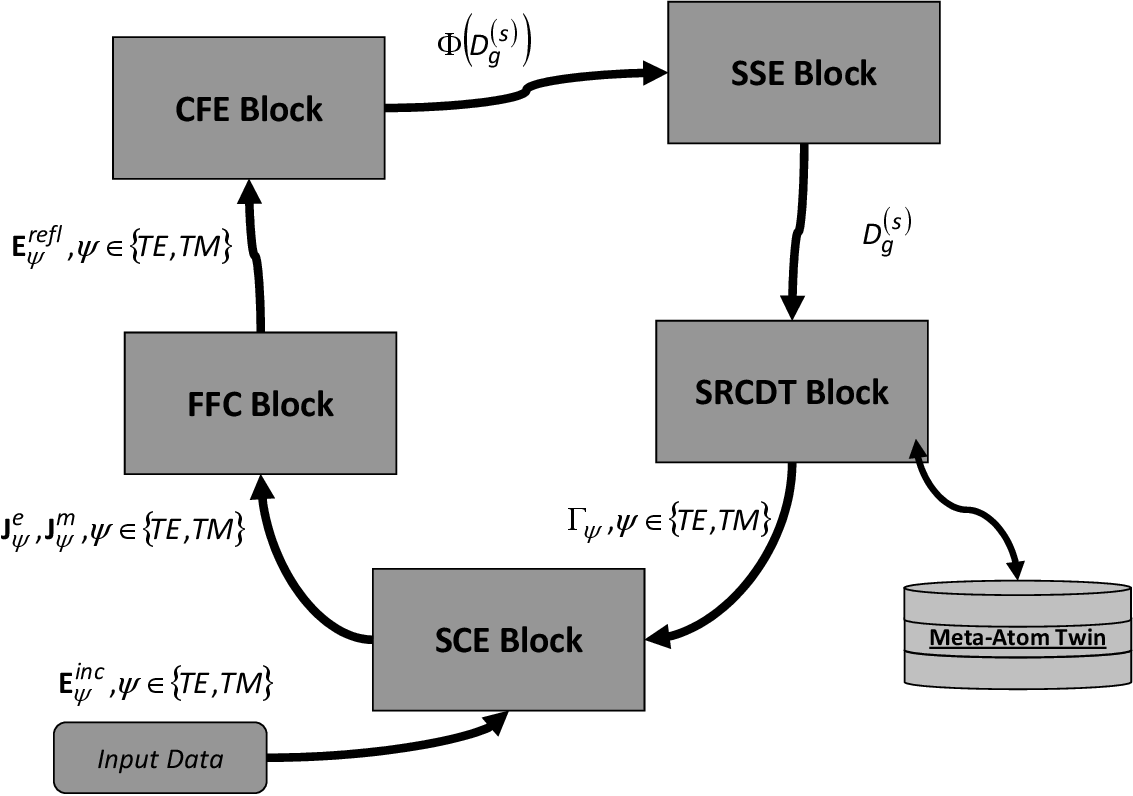}\end{center}

\begin{center}~\vfill\end{center}

\begin{center}\textbf{Fig. 2 - G. Oliveri} \textbf{\emph{et al.,}}
{}``Multi-Functional Polarization-Based Coverage Control through
...''\end{center}

\newpage
\begin{center}~\vfill\end{center}

\begin{center}\begin{tabular}{cc}
\includegraphics[%
  width=0.45\columnwidth,
  keepaspectratio]{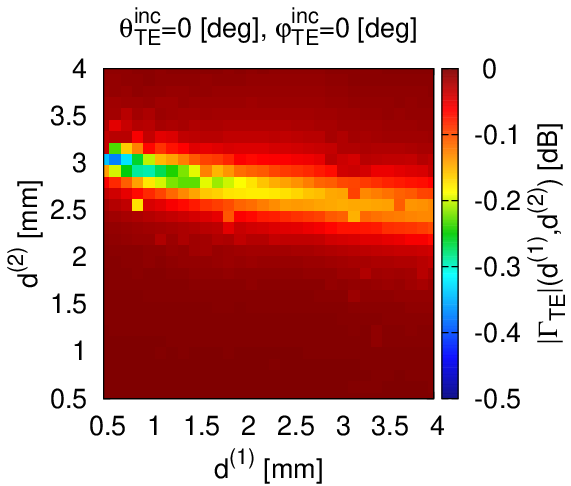}&
\includegraphics[%
  width=0.45\columnwidth,
  keepaspectratio]{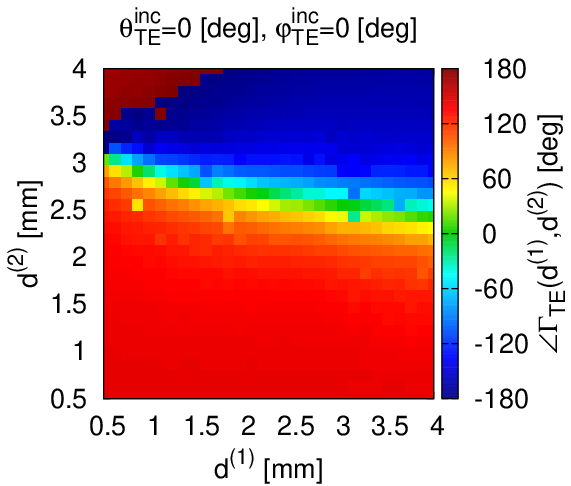}\tabularnewline
(\emph{a})&
(\emph{b})\tabularnewline
\includegraphics[%
  width=0.45\columnwidth,
  keepaspectratio]{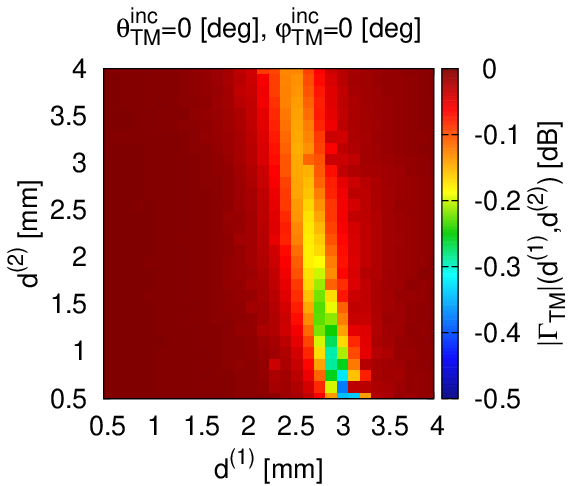}&
\includegraphics[%
  width=0.45\columnwidth,
  keepaspectratio]{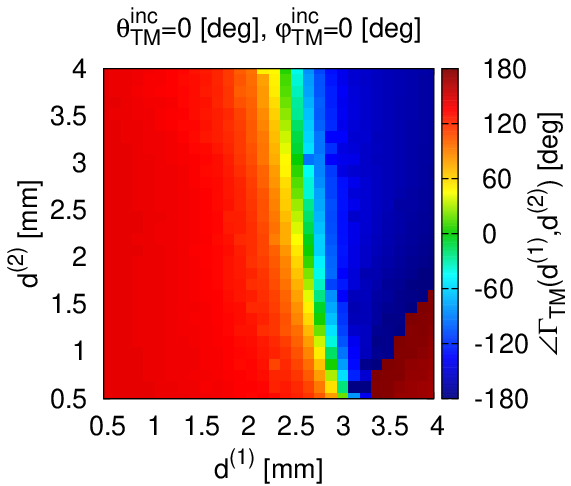}\tabularnewline
(\emph{c})&
(\emph{d})\tabularnewline
\end{tabular}\end{center}

\begin{center}~\vfill\end{center}

\begin{center}\textbf{Fig. 3 - G. Oliveri} \textbf{\emph{et al.,}}
{}``Multi-Functional Polarization-Based Coverage Control through
...''\end{center}

\newpage
\begin{center}~\vfill\end{center}

\begin{center}\includegraphics[%
  width=0.90\columnwidth,
  keepaspectratio]{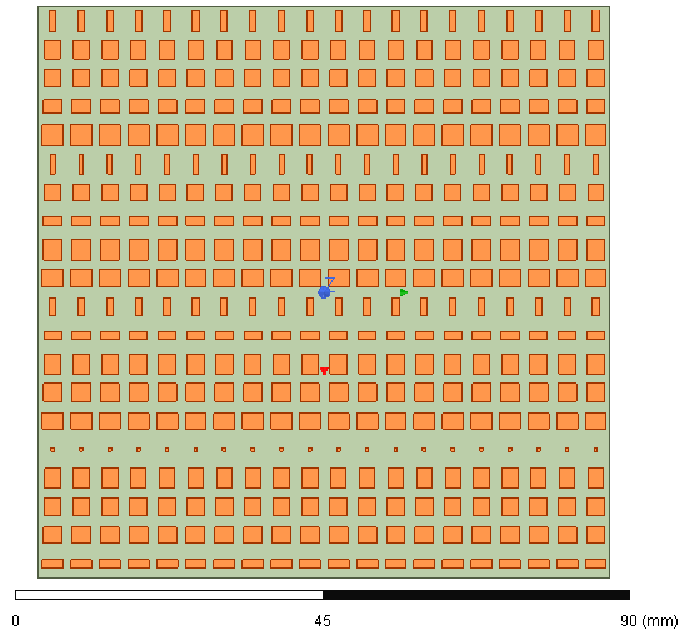}\end{center}

\begin{center}~\vfill\end{center}

\begin{center}\textbf{Fig. 4 - G. Oliveri} \textbf{\emph{et al.,}}
{}``Multi-Functional Polarization-Based Coverage Control through
...''\end{center}

\newpage
\begin{center}~\vfill\end{center}

\begin{center}\begin{tabular}{cc}
\includegraphics[%
  width=0.45\columnwidth,
  keepaspectratio]{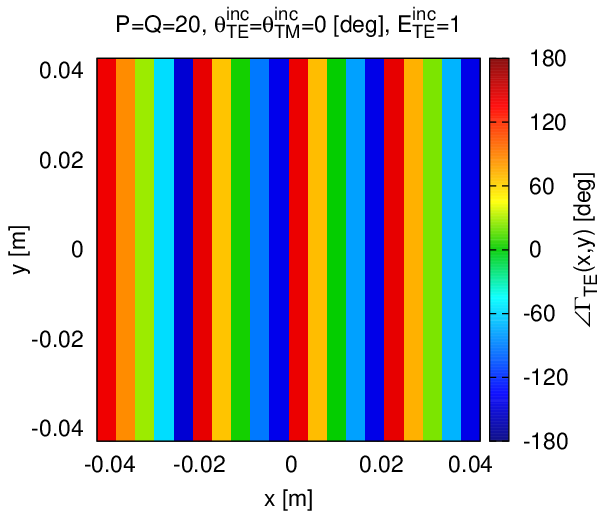}&
\includegraphics[%
  width=0.45\columnwidth,
  keepaspectratio]{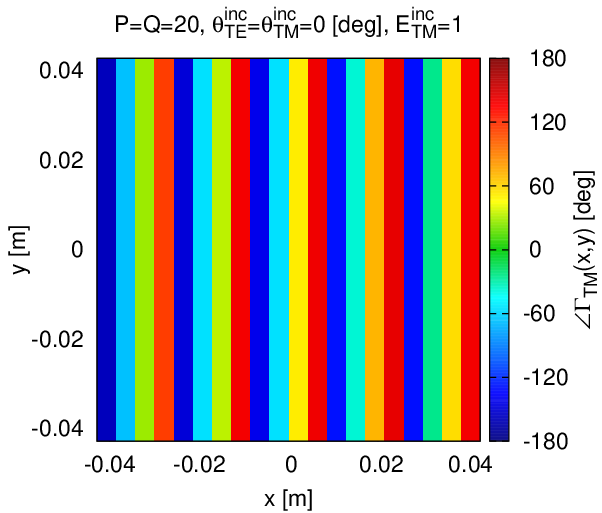}\tabularnewline
(\emph{a})&
(\emph{b})\tabularnewline
\includegraphics[%
  width=0.45\columnwidth,
  keepaspectratio]{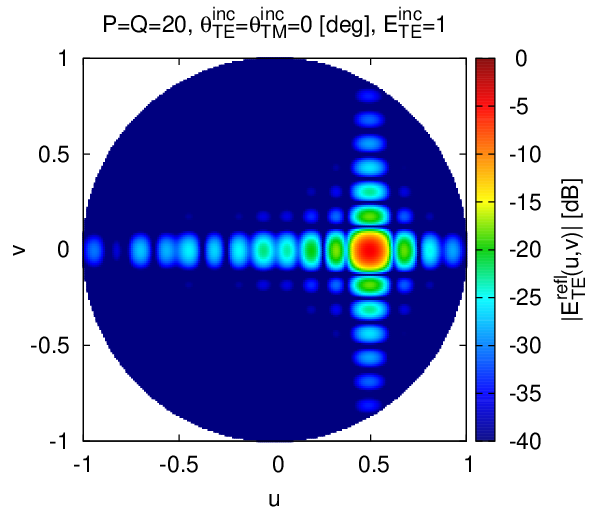}&
\includegraphics[%
  width=0.45\columnwidth,
  keepaspectratio]{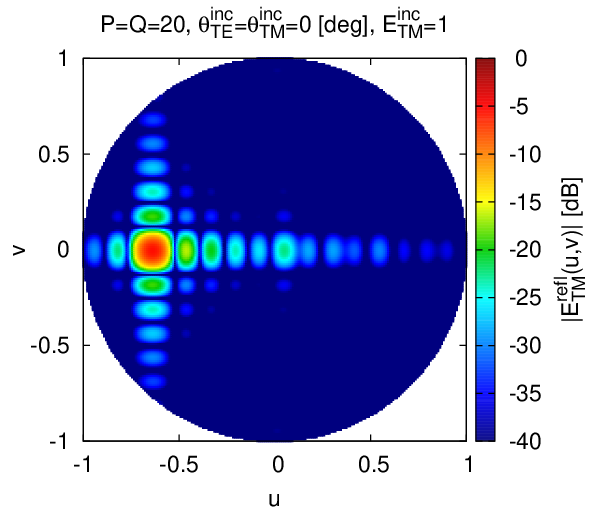}\tabularnewline
(\emph{c})&
(\emph{d})\tabularnewline
\end{tabular}\end{center}

\begin{center}~\vfill\end{center}

\begin{center}\textbf{Fig. 5 - G. Oliveri} \textbf{\emph{et al.,}}
{}``Multi-Functional Polarization-Based Coverage Control through
...''\end{center}

\newpage
\begin{center}~\vfill\end{center}

\begin{center}\includegraphics[%
  width=0.90\columnwidth,
  keepaspectratio]{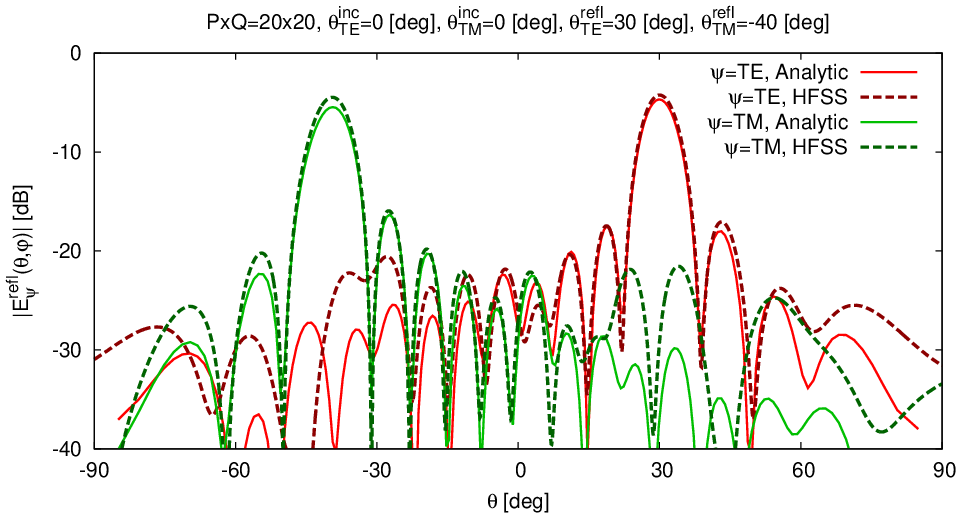}\end{center}

\begin{center}~\vfill\end{center}

\begin{center}\textbf{Fig. 6 - G. Oliveri} \textbf{\emph{et al.,}}
{}``Multi-Functional Polarization-Based Coverage Control through
...''\end{center}

\newpage
\begin{center}~\end{center}

\begin{center}\vfill\end{center}

\begin{center}\begin{tabular}{c}
\includegraphics[%
  width=0.60\columnwidth,
  keepaspectratio]{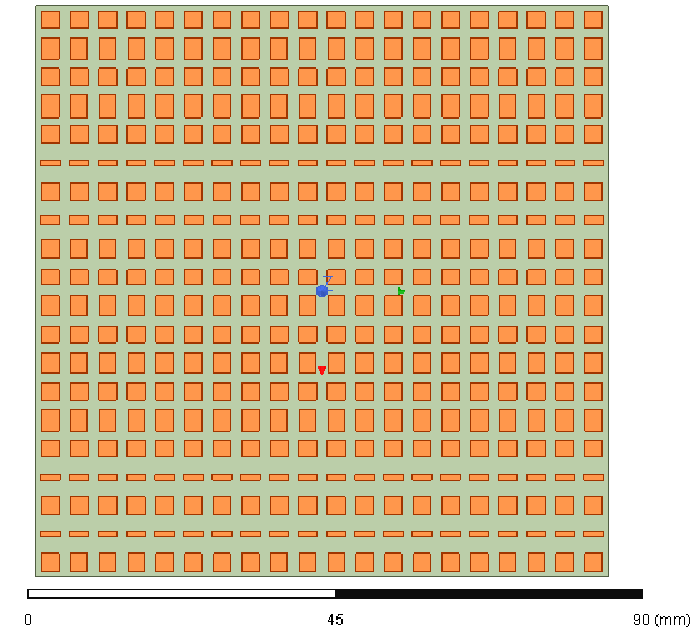}\tabularnewline
(\emph{a})\tabularnewline
\includegraphics[%
  width=0.90\columnwidth,
  keepaspectratio]{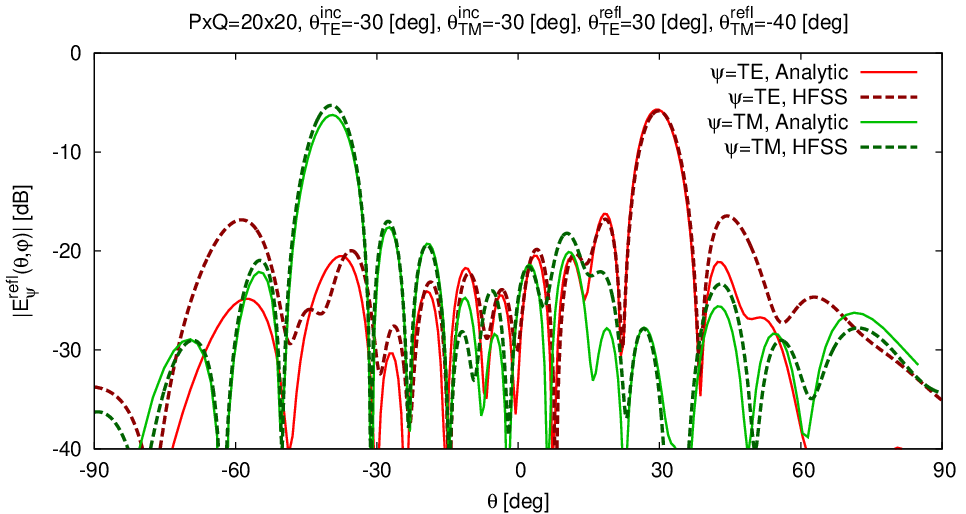}\tabularnewline
(\emph{b})\tabularnewline
\end{tabular}\end{center}

\begin{center}~\vfill\end{center}

\begin{center}\textbf{Fig. 7 - G. Oliveri} \textbf{\emph{et al.,}}
{}``Multi-Functional Polarization-Based Coverage Control through
...''\end{center}

\newpage
\begin{center}~\vfill\end{center}

\begin{center}\begin{tabular}{c}
\includegraphics[%
  width=0.55\columnwidth,
  keepaspectratio]{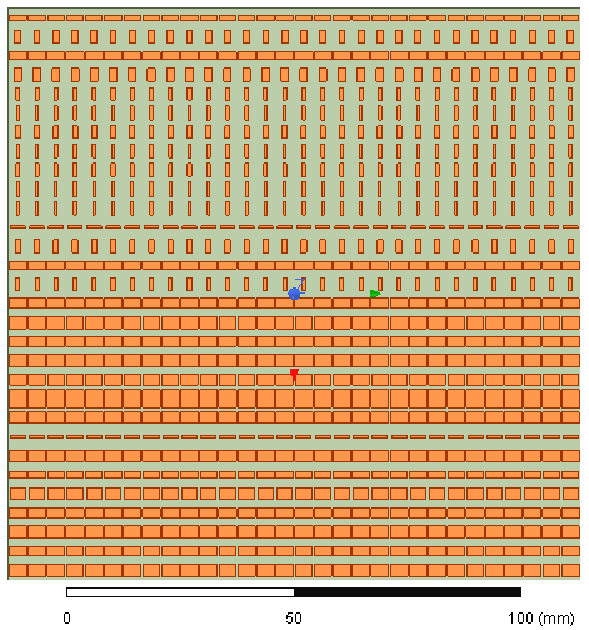}\tabularnewline
(\emph{a})\tabularnewline
\includegraphics[%
  width=0.75\columnwidth,
  keepaspectratio]{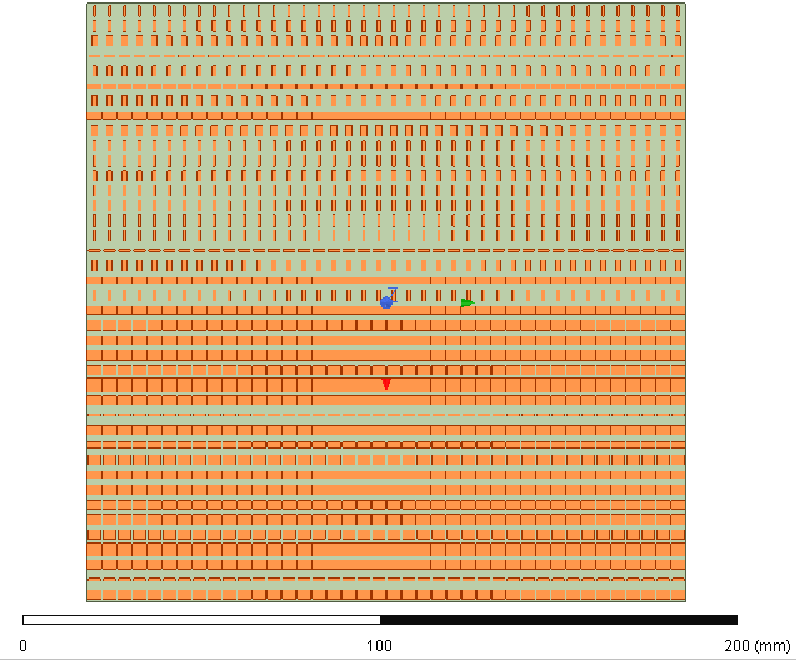}\tabularnewline
(\emph{b})\tabularnewline
\end{tabular}\end{center}

\begin{center}~\vfill\end{center}

\begin{center}\textbf{Fig. 8 - G. Oliveri} \textbf{\emph{et al.,}}
{}``Multi-Functional Polarization-Based Coverage Control through
...''\end{center}

\newpage
\begin{center}~\vfill\end{center}

\begin{center}\begin{tabular}{c}
\includegraphics[%
  width=0.90\columnwidth,
  keepaspectratio]{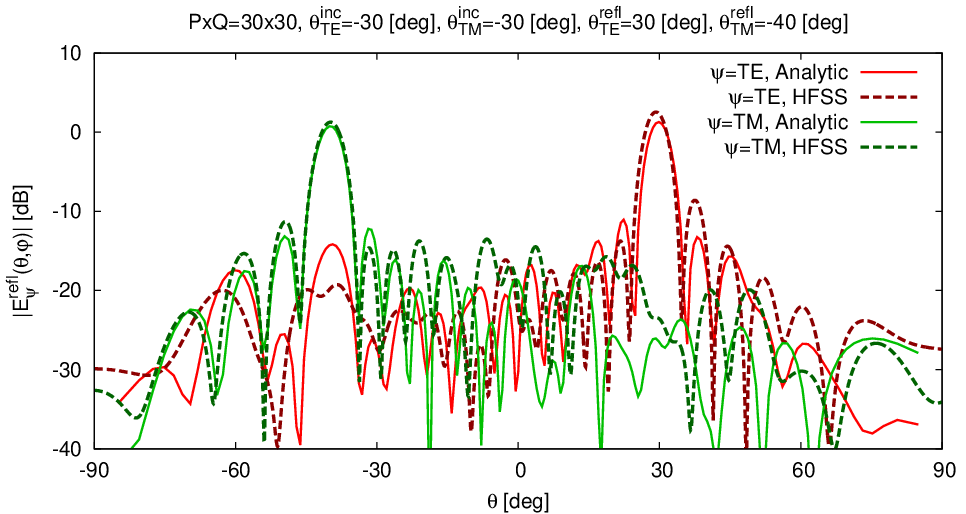}\tabularnewline
(\emph{a})\tabularnewline
\includegraphics[%
  width=0.90\columnwidth,
  keepaspectratio]{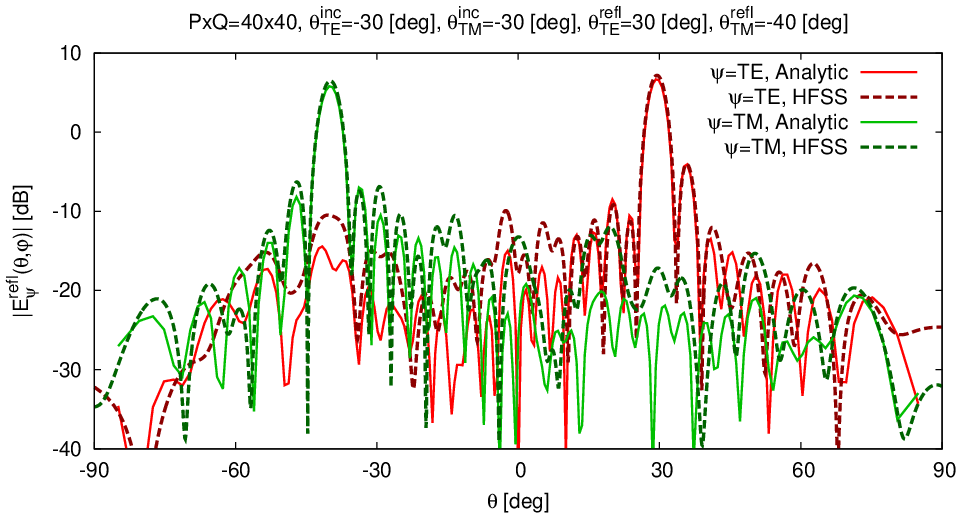}\tabularnewline
(\emph{b})\tabularnewline
\end{tabular}\end{center}

\begin{center}~\vfill\end{center}

\begin{center}\textbf{Fig. 9 - G. Oliveri} \textbf{\emph{et al.,}}
{}``Multi-Functional Polarization-Based Coverage Control through
...''\end{center}

\newpage
\begin{center}~\vfill\end{center}

\begin{center}\begin{tabular}{c}
\includegraphics[%
  width=0.55\columnwidth,
  keepaspectratio]{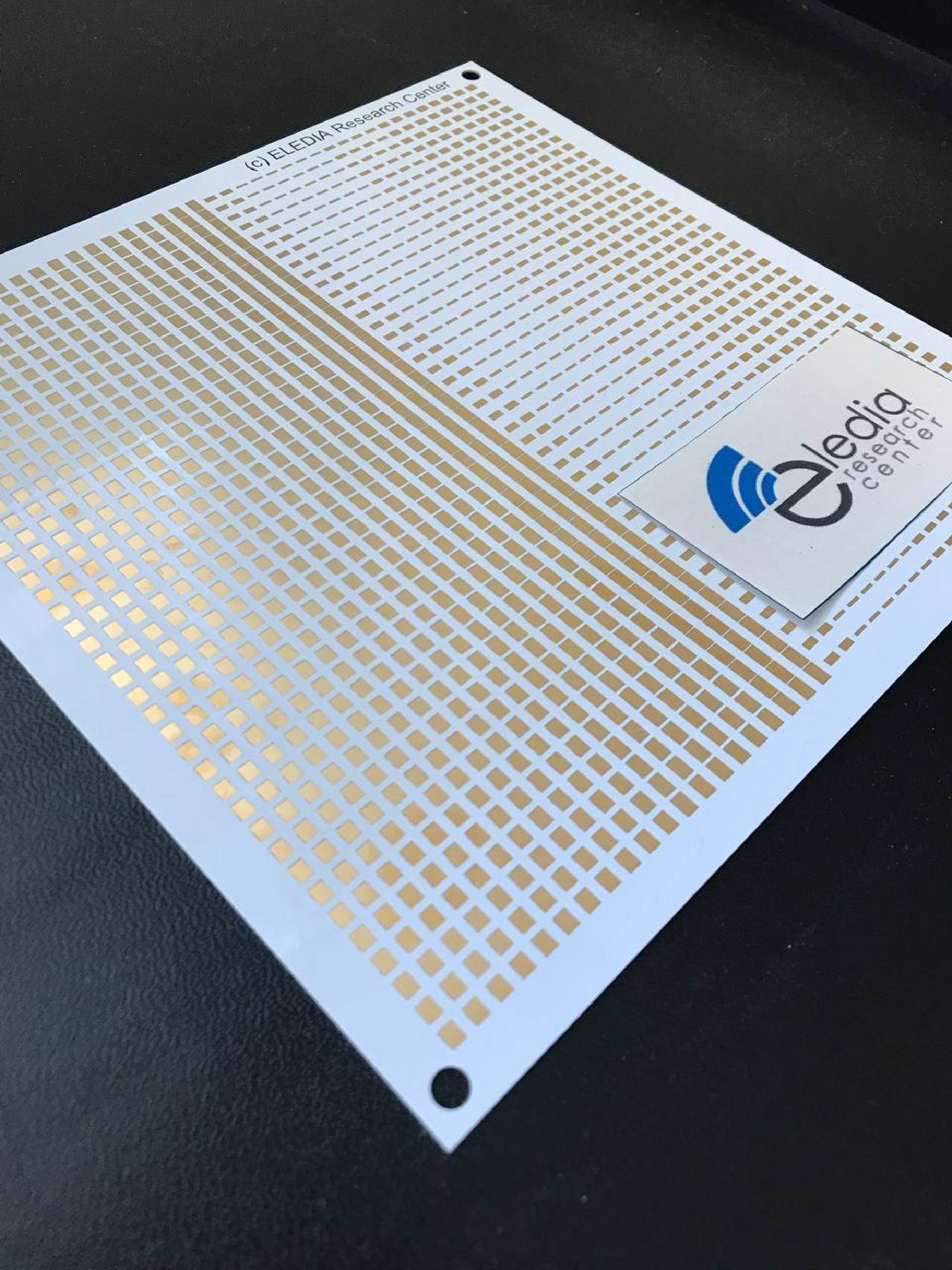}\tabularnewline
(\emph{a})\tabularnewline
\includegraphics[%
  width=0.40\columnwidth,
  keepaspectratio]{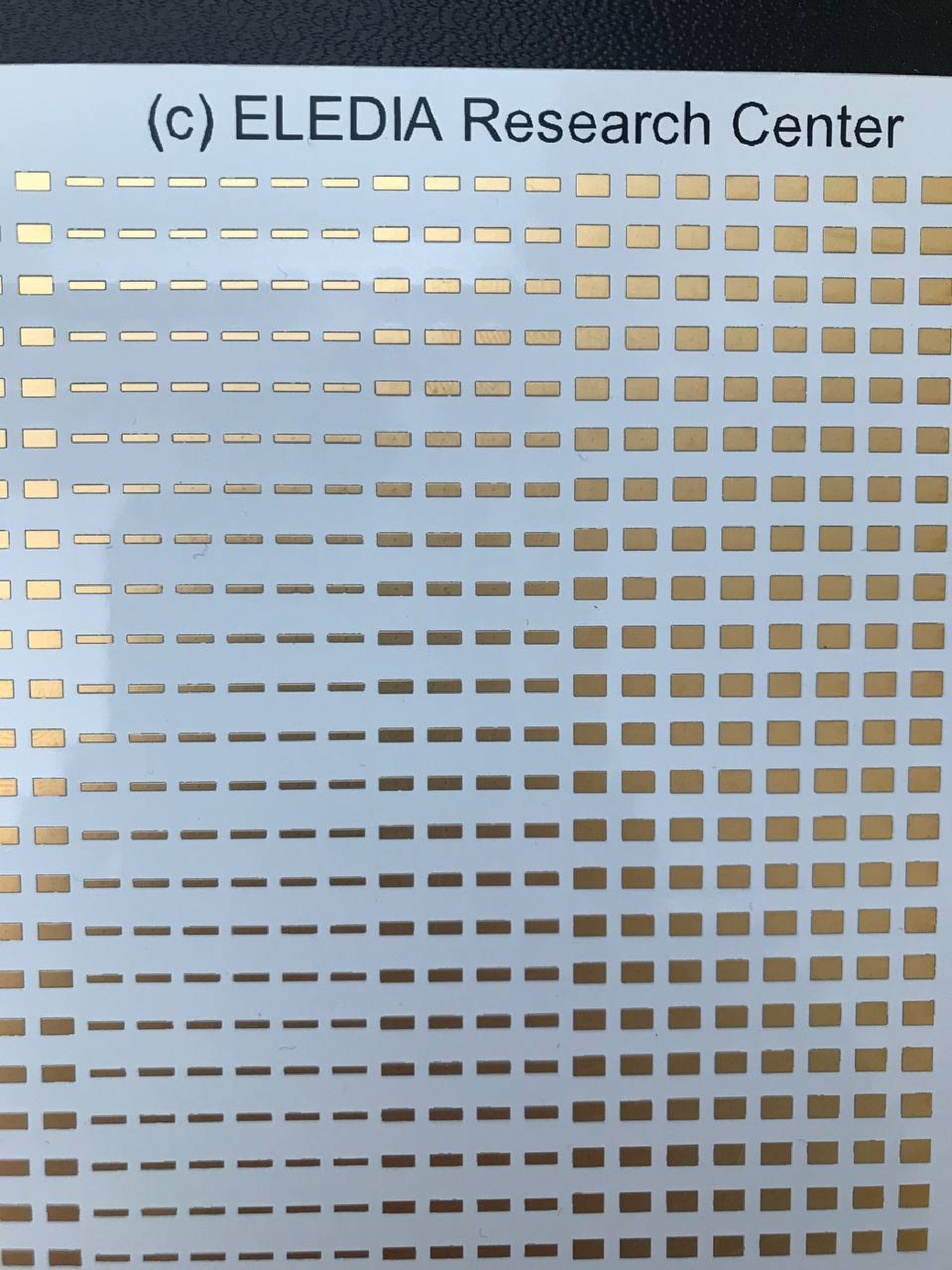}\tabularnewline
(\emph{b})\tabularnewline
\end{tabular}\end{center}

\begin{center}~\vfill\end{center}

\begin{center}\textbf{Fig. 10 - G. Oliveri} \textbf{\emph{et al.,}}
{}``Multi-Functional Polarization-Based Coverage Control through
...''\end{center}

\newpage
\begin{center}~\vfill\end{center}

\begin{center}\begin{tabular}{c}
\includegraphics[%
  width=0.95\columnwidth,
  keepaspectratio]{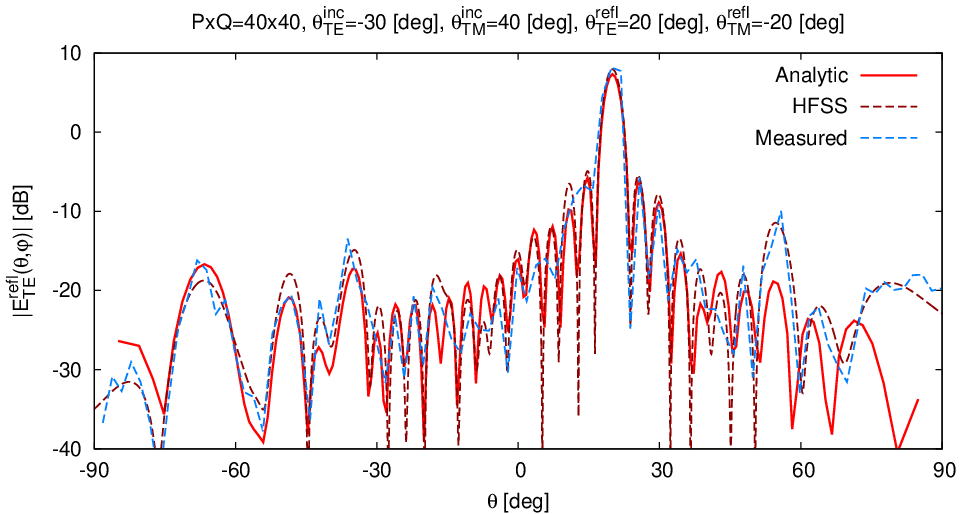}\tabularnewline
(\emph{a})\tabularnewline
\includegraphics[%
  width=0.95\columnwidth,
  keepaspectratio]{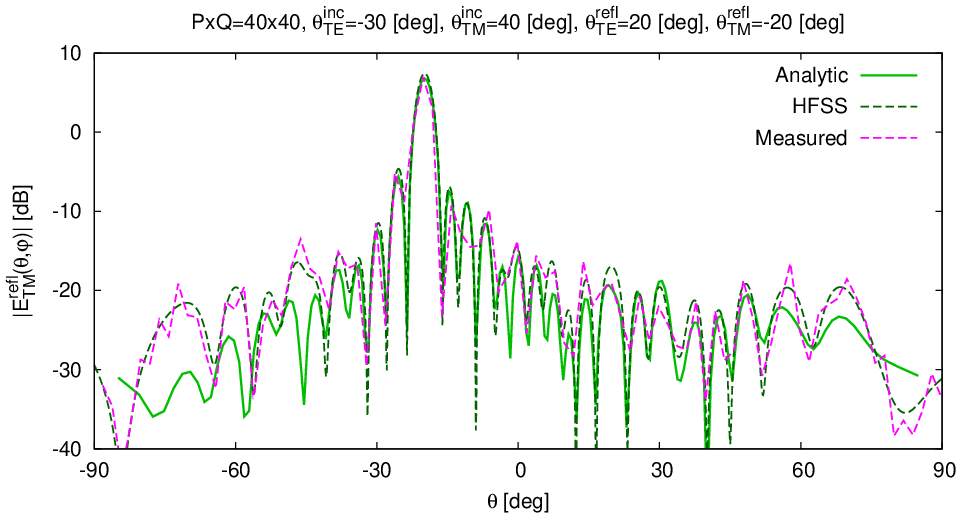}\tabularnewline
(\emph{b})\tabularnewline
\end{tabular}\end{center}

\begin{center}~\vfill\end{center}

\begin{center}\textbf{Fig. 11 - G. Oliveri} \textbf{\emph{et al.,}}
{}``Multi-Functional Polarization-Based Coverage Control through
...''\end{center}

\begin{thebibliography}{10}
\bibitem{Yang 2022}F. Yang, D. Erricolo, and A. Massa, {}``Guest Editorial - Smart Electromagnetic
Environment,'' \emph{IEEE Trans. Antennas Propag}., vol. 70, no.
10, pp. 8687-8690, Oct. 2022.
\bibitem{Massa 2021}A. Massa, A. Benoni, P. Da Ru, S. K. Goudos, B. Li, G. Oliveri, A.
Polo, P. Rocca, and M. Salucci, {}``Designing smart electromagnetic
environments for next-generation wireless communications,'' \emph{Telecom},
vol. 2, no. 2, pp. 213-221, 2021.
\bibitem{Barbuto 2022}M. Barbuto, Z. Hamzavi-Zarghani, M. Longhi, A. Monti, D. Ramaccia,
S. Vellucci, A. Toscano, and F. Bilotti, {}``Metasurfaces 3.0: a
New paradigm for enabling smart electromagnetic environments,'' \emph{IEEE
Trans. Antennas Propag}., vol. 70, no. 10, pp. 8883-8897, Oct. 2022.
\bibitem{Vaquero 2024}A. F. Vaquero, E. Martinez-de-Rioja, M. Arrebola, J. A. Encinar and
M. Achour, {}``Smart electromagnetic skin to enhance near-field coverage
in mm-Wave 5G indoor scenarios,'' \emph{IEEE Trans. Antennas Propag.},
vol. 72, no. 5, pp. 4311-4326, May 2024.
\bibitem{Oliveri 2021c}G. Oliveri, P. Rocca, M. Salucci, and A. Massa, {}``Holographic smart
EM skins for advanced beam power shaping in next generation wireless
environments,'' \emph{IEEE J. Multiscale Multiphys. Comput. Techn.},
vol. 6, pp. 171-182, Oct. 2021.
\bibitem{Benoni 2023}A. Benoni, F. Capra, M. Salucci, and A. Massa, {}``Toward real-world
indoor smart electromagnetic environments - A large-scale experimental
demonstration,'' \emph{IEEE Trans. Antennas Propag}., vol. 71, no.
11, pp. 8450-8463, Nov. 2023.
\bibitem{An 2024}D. An, S. Chang, M. Hwang, Y. Youn, D. Kim, C. Lee, and W. Hong, {}``Diagnosis
and modification of propagating electromagnetic waves using DoA systems
and EM skins,'' \emph{IEEE Trans. Antennas Propag}., vol. 72, no.
4, pp. 3629-3640, Apr. 2024.
\bibitem{Liu 2024}Y. Liu, Z. Liu, S. V. Hum, and C. D. Sarris, {}``An equivalence principle-based
hybrid method for propagation modeling in radio environments with
reconfigurable intelligent surfaces,'' \emph{IEEE Trans. Antennas
Propag}., vol. 72, no. 7, pp. 5961-5973, Jul. 2024.
\bibitem{Yang 2019}F. Yang and Y. Rahmat-Samii, \emph{Surface Electromagnetics with Applications
in Antenna, Microwave, and Optical Engineering}. Cambridge, UK: Cambridge
University Press, 2019.
\bibitem{Alu 2024}A. Alu, N. Engheta, A. Massa, and G. Oliveri, Eds., \emph{Metamaterials-by-Design:
Theory, Technologies, and Vision}. Amsterdam, NL: Elsevier, 2024.
\bibitem{Oliveri 2023c}G. Oliveri, M. Salucci, and A. Massa, {}``Features and potentialities
of static passive EM skins for NLOS specular wireless links,'' \emph{IEEE
Trans. Antennas Propag}., vol. 71, no. 10, pp. 8048-8060, Oct. 2023.
\bibitem{Oliveri 2024}G. Oliveri, F. Zardi, G. Gottardi, and A. Massa, {}``Optically-transparent
EM skins for Outdoor-to-Indoor mm-Wave wireless communications,''
\emph{IEEE Access}, vol. 12, pp. 65178-65191, 2024.
\bibitem{Achouri 2015}K. Achouri, M. A. Salem, and C. Caloz, {}``General metasurface synthesis
based on susceptibility tensors,'' \emph{IEEE Trans. Antennas Propag},
vol. 63, no. 7, pp. 2977-2991, Jul. 2015.
\bibitem{Oliveri 2022b}G. Oliveri, F. Zardi, P. Rocca, M. Salucci, and A. Massa, {}``Building
a smart EM environment - AI-enhanced aperiodic micro-scale design
of passive EM skins,'' \emph{IEEE Trans. Antennas Propag}., vol.
70, no. 10, pp. 8757-8770, Oct. 2022.
\bibitem{Oliveri 2023b}G. Oliveri, M. Salucci, and A. Massa, {}``Generalized analysis and
unified design of EM skins,'' \emph{IEEE Trans. Antennas Propag}.,
vol. 71, no. 8, pp. 6579-6592, Aug. 2023.
\bibitem{Lindell 2019}I. V. Lindell and A. Sihvola, \emph{Boundary Conditions in Electromagnetics}.
IEEE Press, 2019.
\bibitem{Osipov 2017}A. Osipov and S. Tretyakov, \emph{Modern electromagnetic scattering
theory with applications.} John Wiley \& Sons, 2017.
\bibitem{Salucci 2018c}M. Salucci, L. Tenuti, G. Oliveri, and A. Massa, {}``Efficient prediction
of the EM response of reflectarray antenna elements by an advanced
statistical learning method,'' \emph{IEEE Trans. Antennas Propag}.,
vol. 66, no. 8, pp. 3995-4007, Aug. 2018.
\bibitem{Yu 2011}N. Yu, P. Genevet, M. A. Kats, F. Aieta, J. P. Tetienne, F. Capasso,
and Z. Gaburro, {}``Light propagation with phase discontinuities:
generalized laws of reflection and refraction,'' \emph{Science},
vol. 334, no. 6054, pp. 333-337, Oct. 2011.
\bibitem{Rocca 2009w}P. Rocca, M. Benedetti, M. Donelli, D. Franceschini, and A. Massa,
{}``Evolutionary optimization as applied to inverse problems,''
\emph{Inv. Probl}., vol. 25, art no. 123003, pp. 1-41, Dec. 2009.
\bibitem{Oliveri 2022c}G. Oliveri, M. Salucci, and A. Massa, {}``Towards efficient reflectarray
digital twins - An EM-driven machine learning perspective,'' \emph{IEEE
Trans. Antennas Propag}., vol. 70, no. 7, pp. 5078-5093, Jul. 2022.
\bibitem{HFSS 2021}ANSYS Electromagnetics Suite - HFSS (2021). ANSYS, Inc.
\end{thebibliography}
\end{document}